\documentclass[aps,pre,twocolumn,superscriptaddress,showpacs]{revtex4-1}

\usepackage{amsmath}

\usepackage{amssymb}
\usepackage{graphicx}
\usepackage{color}
\usepackage{bm}
\usepackage{color}

\newcommand{\be}{\begin{equation}}
\newcommand{\ee}{\end{equation}}

\newcommand{\benn}{\begin{equation*}}
\newcommand{\eenn}{\end{equation*}}

\newcommand{\bea}{\begin{eqnarray}}
\newcommand{\eea}{\end{eqnarray}}

\newcommand{\mP}{\mathcal{P}}

\newcommand{\mL}{\mathcal{L}}

\begin{document}

\title{First-order condensation transition in the position distribution of a run-and-tumble particle in one dimension}

\author{Francesco Mori}
\affiliation{LPTMS, CNRS, Univ. Paris-Sud, Universit\'e Paris-Saclay, 91405 Orsay, France} 

\author{Giacomo Gradenigo}
\affiliation{Gran Sasso Science Institute, Viale F. Crispi 7, 67100 L’Aquila, Italy}
\affiliation{INFN-Laboratori Nazionali del Gran Sasso, Via G. Acitelli 22, 67100 Assergi (AQ), Italy}

\author{Satya N. Majumdar}
\affiliation{LPTMS, CNRS, Univ. Paris-Sud, Universit\'e Paris-Saclay, 91405 Orsay, France}
\date{\today}
\begin{abstract}
  We consider a single run-and-tumble particle (RTP) moving in one
  dimension. We assume that the velocity of the particle is drawn
  independently at each tumbling from a zero-mean Gaussian
  distribution and that the run times are exponentially distributed. We investigate the probability distribution $P(X,N)$
  of the position $X$ of the particle after $N$ runs, with $N\gg 1$.
  %%
  %% We show that in the regime where $X\sim O(N^{3/4})$, the
  %% distribution of $X$ admits a large deviation form. The corresponding
  %% rate function exhibits a first-order singularity at the critical
  %% value $X=X_c>0$.
  %%
  %%
  We show that in the regime $ X \sim N^{3/4}$ the
  distribution $P(X,N)$ has a large deviation form with a rate
  function characterized by a discontinuous derivative at the critical
  value $X=X_c>0$. The same is true for $X=-X_c$ due to the symmetry
  of $P(X,N)$. We show that this singularity corresponds to a
    first-order condensation transition: for $X>X_c$ a single large
    jump dominates the RTP trajectory. We consider the participation
    ratio of the single-run displacements as the order parameter of the
    system, showing that this quantity is discontinuous at
    $X=X_c$. Our results are supported by numerical simulations
    performed with a constrained Markov chain Monte Carlo algorithm.
  %%with precision smaller than $10^{-100}$.
\end{abstract}

\maketitle
%\tableofcontents

%%%%%%%%%%%%%%%%%%%%%%%%%%%%%%%%%%%%%%%%%%%
%%%%%%%%%%%%%%%%% INTRO %%%%%%%%%%%%%%%%%%%%%
%%%%%%%%%%%%%%%%%%%%%%

\section{Introduction}
\label{sec:intro}

Active systems, characterized by the ability to convert energy from the environment into persistent motion, are ubiquitous in nature. Examples of active matter include flocking of birds \cite{R10,VCB95,VZ12}, active gels \cite{R10,NVG19} and self-propelled bacteria \cite{TC08,C12,berg_book}. The persistent motion of their components drives these systems out-of-equilibrium, giving rise to a wide range of fascinating phenomena. Even though several of these features arise from the complex interactions of several components \cite{TC08,C12,CT15,BDL16}, many interesting phenomena, e.g., the universality of the survival probability \cite{MLDM20a,MLDM20,DMS21}, can be already observed at the single-particle level, where one can often obtain exact analytical results.

One of the most studied models of active matter is the run-and-tumble particle (RTP). This model was originally known as persistent random walk \cite{kac74,S87,Orshinger90,W02,HJ95,ML17} and has been applied in recent years to describe the directed motion of a class of bacteria, including \emph{E. Coli} \cite{berg_book,C12,TC08,CT15,SFB15}. These bacteria typically move alternating between running phases of straight motion with constant velocity, to tumblings, i.e., sudden changes of direction. Despite its apparent simplicity, this model encapsulates several general features of active matter, including motility-induced phase separation \cite{CT15} and non-Boltzmann steady states in the presence of a confining potential \cite{TC08,DKM18,SAC19,MBE19,BMR20}.

\begin{figure}
\includegraphics[width=\columnwidth]{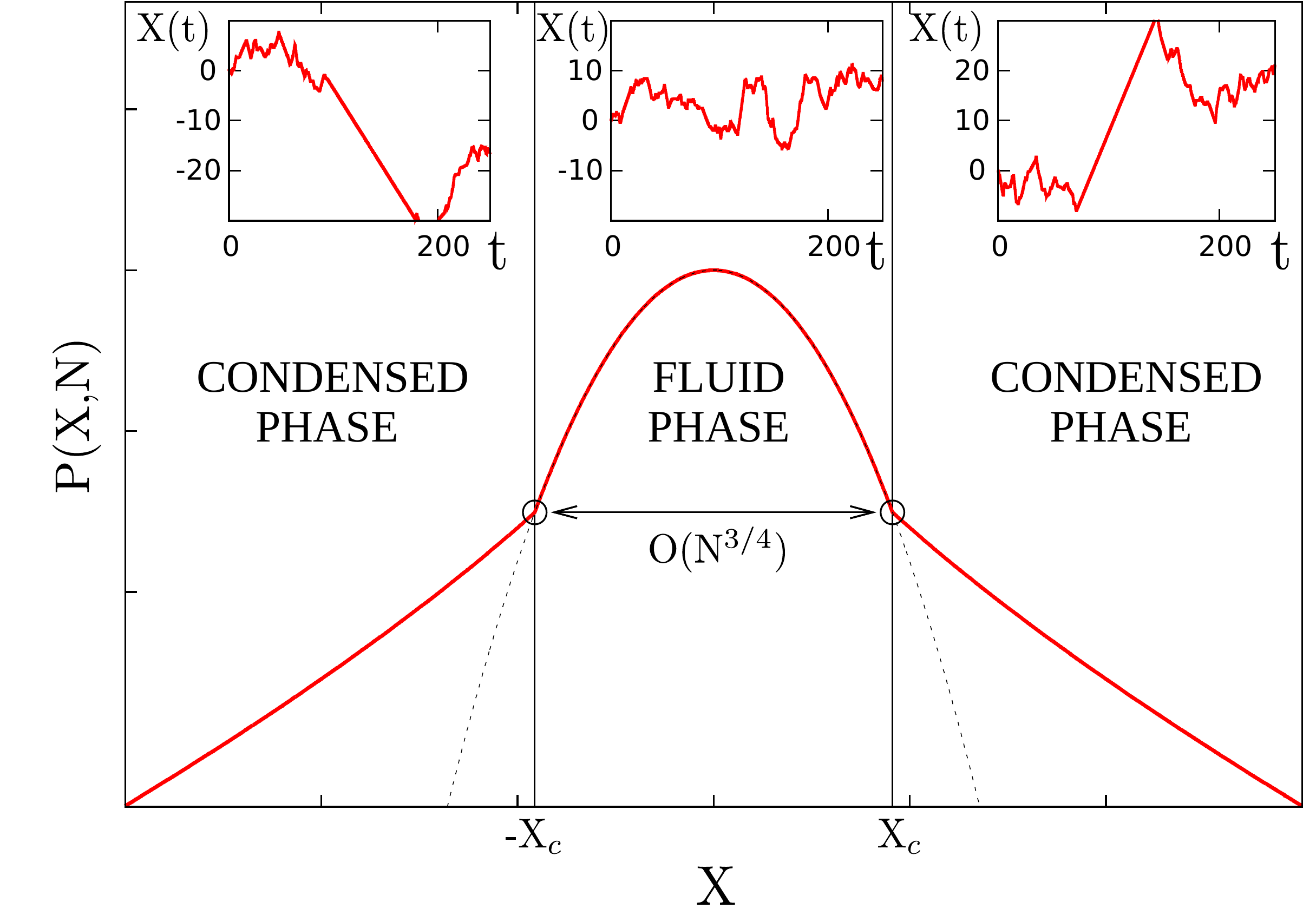}
\caption{\emph{Main:} Schematic representation of the PDF $P(X,N)$ of the final position $X$ of a single run-and-tumble particle (RTP) after $N\gg 1$ running phases. The central region $-X_c<X<X_c$ [with $X_c\sim O(N^{3/4})$] describes the fluid phase where $P(X,N)$ has a Gaussian shape. For $X>X_c$ (and by symmetry for $X<-X_c$) the system is in the condensed phase, where $P(X,N)$ has an anomalous shape. \emph{Insets}: Typical RTP trajectories $X(t)$ as a function of the time $t$. In the condensed phase (external panels), the trajectory is dominated by a single run. In the fluid phase (central panel), the runs contribute to the total displacement by roughly the same amount.}
\label{fig1}
\end{figure}

One of the simplest and most natural observables that one can investigate for the RTP model in one dimension is the probability density function (PDF) $P(X,N)$ of the position $X$ of a single RTP after $N$ running phases \cite{S87,MAD12,DM12,GM19,SBS20,PTV2020,MLDM21}. For the RTP model and its many variants, computing $P(X,N)$ for any $N$ is usually nontrivial. Note that, if the initial velocity is chosen at random, the PDF $P(X,N)$ is symmetric around $X=0$, i.e. $P(X,N)=P(-X,N)$. For $N\gg 1$, as a consequence of the central limit theorem (CLT), one expects $P(X,N)$ to be Gaussian in the region $|X|\sim \sqrt{N}$ \cite{C12}. However, outside the range of validity of the CLT, i.e., for $|X|\gg \sqrt{N}$, the shape of $P(X,N)$ depends on the details of the model and is usually not Gaussian. The large-deviation tails of $P(X,N)$ have been studied for a large class of RTP models \cite{ DM12,GM19,SBS20,PTV2020,MLDM21,M21}, including RTPs moving in $d$ dimensions (for which $X$ represents the $x$-component of the position of the particle) \cite{PTV2020,MLDM21} and RTP models for which the speed $v$ of the particle during each running phase is drawn from some distribution $W(v)$ \cite{ZSS08,GM19,MLDM21}. In particular, it has been shown that, under certain conditions \cite{MLDM21}, the system undergoes a condensation phase transition at some critical value $X_c$ of the position $X$ \cite{MLDM21}. Below the transition, the different running phases contribute to the total displacement $X$ by roughly the same amount. Conversely, for $|X|>X_c$, a single running phase of size $X_{\rm cond}$, called \emph{condensate}, dominates the trajectory, contributing to a finite fraction of $X$ (see the insets in Fig. \ref{fig1}). This condensation transition leads to a non-analytic behavior of the PDF $P(X,N)$ at the critical points $X=\pm X_c$ (see Fig. \ref{fig1}).

In the vicinity of the transition, the PDF $P(X,N)$ can be written as
\begin{equation}
P(X,N)\simeq \exp\left[-N^{2\alpha-1} F\left(\frac{X}{N^\alpha}\right)\right]\,,
\label{definiton_F}
\end{equation}
where $F(z)$ is the rate function associated to the large-deviation regime and the exponent $\alpha>0$, that depends on the model, determines the scale of the large deviations. Notably, the dynamical phase
transition is signaled by the non-analyticity of $F(z)$ at the
critical point $z=z_c$, where $z_c=X_c/N^{\alpha}$. In particular, the
transition is said to be of order $n$ if the $n$-th derivative of
$F(z)$ is discontinuous at $z=z_c$. For instance, in
Ref. \cite{MLDM21}, it was shown that for an RTP in $d$ dimensions and
for a family on the speed distributions $W(v)$ the system undergoes a
transition of order $n\geq 2$, where $n$ depends on the system
parameters. Similarly, in Ref. \cite{GM19} a one-dimensional RTP in
the presence of a constant drive $E>0$ and with Gaussian speed
distribution $W(v)$ was considered. In this case, it was shown that
the system undergoes a first-order phase transition, i.e.,
  $n=1$. However, it was not clear whether a first-order
transition could be observed for an RTP system without an
external drive. Moreover, it is also relevant to ask whether or not
this first-order dynamical phase transition is associated with a
discontinuity of some order parameter, as observed for equilibrium phase transitions.

In addition to RTP systems, this kind of real-space condensation
transitions has been observed in a wide range of situations
\cite{EH05,M2008}.  Examples include the discrete nonlinear
Schr\"odinger equation~\cite{RCK2000,SPP17,GIL21,GIL21b,GIP21}, economic and
financial models \cite{BM2000,BJJ2002,FZV13}, and mass-transport
models
\cite{MEZ2005,EMZ06,EHM06,EM08,EMPT10,SEM2014,SEM2014b,SEM2016,GB2017}.
In these systems, a phase transition is observed when a control
parameter, e.g., the total mass of the system~\cite{MEZ2005}
or the total energy~\cite{GIL21}, is increased above a critical
threshold. Above this critical point, a condensate appears in real
space absorbing a macroscopic fraction of the total mass. For
instance, in the context of mass-transport models on lattices the
condensate is a single lattice site carrying a finite fraction of the
total mass. Similarly, in the context of wealth distribution in a
population, the analogous of a condensate would be an extremely
wealthy individual. In the case of RTPs the condensate is a single
running phase which dominates the trajectory.

In this paper, we consider a single RTP on a line. We choose the velocity distribution $W(v)$ to be Gaussian and the distribution of the time between two tumblings to be exponential. We investigate the
distribution $P(X,N)$ of the position of the particle in the late-time
limit. We show that in the large-deviation regime where $X\sim
N^{3/4}$, corresponding to $\alpha=3/4$ in Eq. \eqref{definiton_F}, the particle undergoes a first-order phase transition and we
compute exactly the corresponding rate function $F(z)$. Moreover, we
provide a detailed description of the mechanism of this transition. Above the transition a condensate, i.e., a single displacement of length $X_{\rm cond}\sim N^{3/4}$, appears. We
identify the relevant order parameter for the system, showing that it
undergoes a jump discontinuity at $z=z_c$. We also verify our
results by performing high-precision numerical simulations. This RTP
model corresponds to the one considered in Ref.~\cite{GM19} but with
no external drive ($E=0$). Note that $E=0$ is a singular point. This is because several features associated with the condensation transition for $E>0$ \cite{GM19} are different from the $E=0$ case. First, in the case $E>0$, the left and right tails of the PDF $P(X,N)$ are not symmetric, due to the finite drive. Moreover, for $E>0$, the system undergoes a phase transition at the critical point $X_c=EN+bN^{2/3}$, where $b$ is some constant of order one. Above the transition a condensate of size $X_{\rm cond}\sim N^{2/3}$ appears. Thus the condensate displacement $X_{\rm cond}$ is subleading with respect to the total displacement $X$. On the other hand, for $E=0$ the PDF $P(X,N)$ is symmetric around $X=0$, i.e., $P(X,N)=P(-X,N)$. Furthermore, the transition occurs when $X$ exceeds the critical value $X_c=z_c N^{3/4}$ and above the transition the condensate size $X_{\rm cond}$ scales as $N^{3/4}$. Thus, in the case $E=0$ the condensate size $X_{\rm cond}$ is of the same order as the total displacement $X$.

The rest of the paper is organized as follows. In Section
\ref{sec:model-results} we provide the details of the model and we
present a summary of the salient results. The details of the
computation of the late-time position distribution of the RTP are
presented in Sec. \ref{sec:position}. In Section \ref{sec:marginal},
we investigate the marginal probability distribution of the
displacement of the RTP during a single running phase and we identify
the relevant order parameter of the system. Finally, in Section
\ref{sec:conclusions} we conclude with a summary of the paper and few
remarks. Some details of the computations and of the numerical simulations are presented in the
Appendices.

\section{The model and the summary of the main results}
\label{sec:model-results}

\subsection{ The model}

We consider a single RTP on a line, starting initially at the origin. The particle chooses a velocity $v_1$ (positive or negative), drawn from the distribution $W(v)$, and starts to move with constant velocity $v_1$. After some time $\tau_1$, drawn from the time distribution $p(\tau)$, the particle tumbles, i.e., it chooses a new velocity $v_2$, independently drawn from $W(v)$. Then, it starts moving with the new velocity $v_2$, until it tumbles again after some random time $\tau_2$, drawn from $p(\tau)$. We assume that the tumblings happen instantaneously and that the running times are drawn from an exponential distribution with average value $1/\gamma$, i.e., that
\be
p(\tau) = \gamma\exp(-\gamma \tau)\theta(\tau)
\label{duration_pdf}
\ee
where $\theta(\tau)$ is the Heaviside theta function, i.e., $\theta(\tau)=1$ for 
$\tau\ge 0$ and $\theta(\tau)=0$ for $\tau<0$. The parameter $\gamma$ is the tumbling rate of the RTP. Moreover, we assume that the velocity distribution is Gaussian with zero mean and variance $\sigma^2$. In other words, we choose
\begin{equation}
W(v) = \frac{1}{\sqrt{2\pi\sigma^2}} \exp\left[ -v^2/(2\sigma^2)\right] \,.
\label{vel_pdf} 
\end{equation}
 For the sake of simplicity, we set $\sigma^2=\gamma=1$ in the rest of the paper.

When considering the RTP model, one can either observe a trajectory up to some fixed time $T$ (\emph{fixed-$T$ ensemble}) or until $N$ running phases are completed (\emph{fixed-$N$ ensemble}). Accordingly, one either studies the PDF $P(X,T)$ of the position $X$ of the RTP after time $T$ or the PDF $P(X,N)$ of $X$ after $N$ running phases. In Ref. \cite{MLDM21} it was shown that the late time behavior is qualitatively similar for the two ensembles. Since performing the computation is technically easier at fixed $N$, for the sake of simplicity we focus here on the fixed-$N$ ensemble. Our results can be generalized to the fixed-$T$ case. The total displacement of the particle after $N$ runs is given by
\be X= \sum_{i=1}^N x_i\,,
\label{eq:Xacc}
\ee
where $x_i=v_i\tau_i$ is the displacement during the $i$-th running phase. The velocities $v_1\,,\ldots\,,v_N$ are independent and identically distributed (i.i.d.) random variables drawn from the PDF in Eq. \eqref{vel_pdf}. Similarly, the times $\tau_1\,,\ldots\,,\tau_N$ are i.i.d. exponentially distributed random variables with rate $\gamma$.

Thus, the distribution of a single-run displacement $x_i$ is given by
\be
\mP(x)= \int_{-\infty}^{\infty} dv \int_0^{\infty}d\tau~ W(v)\,p(\tau)\,
\delta(x- v \tau)\, \,. 
\label{x_marg.1}
\ee
Using the expressions for $p(\tau)$ and $W(v)$ given in Eqs. 
(\ref{duration_pdf}) and (\ref{vel_pdf}) respectively, one obtains 
\be 
\mP(x)=\frac{1}{\sqrt{2\pi}}\int_{0}^{\infty}d\tau~\frac{1}{\tau}e^{-\tau-x^2/(2\tau^2)}\,.
\label{x_marg.2a}
\ee
It turns out that this expression can be written in terms of the Meijer G-function $G_{0,3}^{3,0}(x|\ldots)$, which can be evaluated using Mathematica, as
\be
\mP(x)= \frac{1}{2\sqrt{2}\pi} G_{0,3}^{3,0} \left(\begin{array}{c|ccc} & 0 & 0 & 0 \\ x^2/8 &  &  &  \\  & 0 & 0 & 1/2 \end{array}\right).
\label{x_marg.2}
\ee
From this expression in Eq. \eqref{x_marg.2} we observe that the marginal distribution $\mP(x)$ is symmetric around $x=0$. The mean and the variance of $x=v\tau$ can be easily computed. While the mean is simply 
\be
\langle x \rangle =0\,,
\label{mean_x}
\ee
the variance  reads
\be
\langle x^2\rangle- \langle x\rangle^2 =  \langle v^2\rangle \langle \tau^2\rangle =2\,.
\label{var_x}
\ee
From the expression for the marginal distribution $\mP(x)$ in Eq. \eqref{x_marg.2a} one can show that when $|x|\to \infty$ (see Appendix \ref{app:Asymp_Px})
\be \mP(x) \approx \frac{1 }{\sqrt{3}\,
  |x|^{1/3}}\, e^{-3\, | x|^{2/3}/2}\, .
\label{eq:mP-asymptotic}
\ee
Thus the jump distribution $\mP(x)$ has a stretched exponential tail $\sim 
\exp[-(3/2)|x|^{2/3}]$. Notably, the single-run PDF $\mP(x)$ satisfies the condition for condensation presented in Ref. \cite{MLDM21}. This criterion was derived using a grand canonical argument and states that if the PDF $\mP(x)$ satisfies
\begin{equation}
e^{-c|x|}<\mP(x)<1/|x|^3\,
\label{criterion}
\end{equation}
for large $|x|$, where $c>0$ is any positive constant, then the corresponding RTP model displays a condensation transition. Since the PDF in Eq. \eqref{eq:mP-asymptotic} satisfies the condition in Eq. \eqref{criterion}, we expect that the system undergoes a dynamical phase transition in the large-deviation regime of $X$. Note that the standard RTP model with fixed velocity $v_0$ would correspond to the choice
\begin{equation}
W(v)=\frac12\delta(v-v_0)+\frac12\delta(v+v_0)\,.
\end{equation}
However, using Eq. \eqref{x_marg.1}, it is easy to show that the displacement distribution $\mP(x)=e^{-|x|/v_0}/(2v_0)$ does not satisfy the criterion in Eq. \eqref{criterion} and thus no condensation transition occurs in this case.

Interestingly, the condition in Eq. \eqref{criterion} is satisfied for several other choices of the speed distribution $W(v)$. Indeed, the marginal distribution $\mP(x)$ and the PDF $W(v)$ are related by \eqref{x_marg.1}. For a list of possible distributions $W(v)$ that lead to condensation see Ref. \cite{MLDM21}. Moreover, it is clear from Eq. \eqref{x_marg.1} that the choice of the running-time distribution $p(\tau)$, which is assumed to be exponential in this work, also affects the tail behavior of $\mP(x)$ and could also lead to condensation. Note however that the criterion in Eq. \eqref{criterion} does not provide information on the order of the transition or the scale at which the transition occurs. To determine these features a comprehensive analysis is required.

In Ref. \cite{MLDM21} a detailed analysis of the condensation transition was carried out in the case where $\mP(x)$ has a power-law tail for large $|x|$. The main goal of this paper is instead to investigate the case where $\mP(x)$ has a stretched exponential tail, for which both the scale and the mechanism of the phase transition are different from the power-law case. In particular, in the stretched-exponential case the size of the condensate scales as $N^{3/4}$ and the transition is first-order. Conversely, in the model studied in \cite{MLDM21}, the condensate mass scales linearly in $N$ and the transition is of order $n\geq 2$.

Note that the case where $\mP(x)$ has a stretched exponential tail has also been studied in the context of the discrete nonlinear
Schr\"odinger equation~\cite{GIL21,GIL21b,GIP21}. However, in that case the distribution $\mP(x)$ has support only for positive $x$ and the phase transition occurs at a different scale, namely for $X\sim O(N)$. In this paper instead the variable $x$ can be positive or negative and $\mP(x)$ is symmetric around $x=0$.

Given the marginal distribution $\mP(x)$ in Eq. (\ref{x_marg.1}) for the displacements $x_1\,,\ldots,x_N$, we are interested in computing the PDF $P(X,N)$ of the final position $X=\sum_{i=1}^N x_i$. Since the displacements $x_1\,,\ldots,x_N$ are i.i.d. random variables, their joint probability distribution is simply given by the product of the marginal probabilities, and we obtain
\be
P(X,N) =  \prod_{i=1}^N \int_{-\infty}^{\infty}
dx_i~\mP(x_i)~\delta\left(X-\sum_{i=1}^N x_i\right),
\label{eq:PXN-free}
\ee
where the delta function enforces the final position to be $X$ and $\mP(x)$ is given in Eq. (\ref{x_marg.1}).

It is interesting to notice that one can rewrite Eq. \eqref{eq:PXN-free} as
\be
P(X,N) =  \prod_{i=1}^N \int_{-\infty}^{\infty}
dX_i~\mP(X_i-X_{i-1})~\delta\left(X-X_N\right),
\label{eq:PXN-freeb}
\ee
where we have defined
\be 
X_i=x_1+x_2+\ldots+x_i\,,
\ee
with $X_0=0$. The variable $X_i$ can be interpreted as the position after $i$ steps of a one-dimensional random walker with jump distribution $\mP(x)$. Note that $X_N=X$ is the final position of the walker. Thus,
\be 
P_{\rm traj}(X_0=0,X_1,\ldots,X_N=X)=\prod_{i=1}^N dX_i~\mP(X_i-X_{i-1})\label{Ptraj}
\ee
gives the probability of the trajectory of a discrete-time random walk of $N$ steps. This is a well-studied model with several applications \cite{M10}.

There is yet another interesting interpretation of Eq. \eqref{Ptraj}. We first rewrite Eq. \eqref{Ptraj} as
\be 
P_{\rm traj}(\{X_i\})=e^{-E[\{X_i\}]}\,,\label{Ptraj2}
\ee
with
\begin{equation}
E[\{X_i\}]=-\sum_{i=1}^{N}\log\left[\mP\left(X_i-X_{i+1}\right)\right]\,.
\end{equation}
Then we can interpret $P_{\rm traj}(\{X_i\})$ as the equilibrium Boltzmann measure with $E[\{X_i\}]$ representing the energy of a gas of $N$ particles on a line with position cohordinates $\{X_i\}$ with nearest-neighbor interactions. This model is particularly relevant in the context of $(1+1)$-dimensional solid-on-solid models, where the variable $X_i$ represents the height of a fluctuating interface at the $i$-th site of a substrate of length $N$ \cite{SM06}. Alternatively, $X_i$ could describe the position of the $i$-th monomer in a polymer chain consisting of $N$ monomers. In particular, since the distribution $\mP(x)$ that we consider is non-Gaussian, our model would correspond to non-harmonic interactions between neighboring monomers \cite{PVV17}.

It is clear from Eq. \eqref{eq:PXN-free} that studying the
distribution of $X$ simply amounts to the classical problem of finding
the distribution of the sum of $N$ i.i.d. random variables, each drawn
from the symmetric distribution $\mP(x)$. This problem has been
extensively investigated in the probability literature
\cite{Feller_book} and has recently been studied for correlated variables \cite{HB21}. In particular, the case where $\mP(x)$ has
stretched exponential tails $\mP(x)\sim e^{-a |x|^{\beta}}$, with
$a>0$ and $0<\beta<1$ was first investigated by Nagaev
\cite{nagaev1,nagaev2}, who identified the presence of a nontrivial
large deviation regime of $P(X,N)$ in the region $|X|\sim
N^{1/(2-\beta)}$. Note that our RTP model corresponds to $a=3/2$ and
$\beta=2/3$. Moreover, the rate function associated to this
large-deviation regime was derived for any $\beta>0$ in a recent
mathematical work \cite{BKL20}. In this paper, we present an
alternative derivation of the rate function $F(z)$, which is in
agreement with the result of Ref. \cite{BKL20} (see appendix \ref{app:equivalence}). In addition, we provide a detailed analysis of the mechanism of the phase transition,
investigating the marginal probability distribution of a single-run
displacement and identifying the order parameter associated to the
transition, which were not addressed in Ref. \cite{BKL20}.

\subsection{The summary of the main results}

Since the detailed derivations are somewhat technical, it is useful to provide a summary of the salient features of our main results. This is the purpose of this section, while the detailed derivations are presented in the following sections. We provide a threefold description of the dynamical phase transition, based on the analysis of three main observables: (1) the position distribution of the particle, (2) the marginal probability of a single-run displacement, and (3) the participation ratio, i.e., the order parameter associated to the condensation transition.

\subsubsection{Position distribution}

Our first goal is to investigate the distribution $P(X,N)$ of the position $X$ of the RTP after $N$ running phases. In the limit of large $N$ we identify three distinct regimes. In the typical regime $|X| \sim \sqrt{N}$, the distribution of $X$ is Gaussian,
\begin{equation}
P(X,N)\sim e^{-X^2/(4N)}\,,
\label{gaussian_tail}
\end{equation} 
as a consequence of the CLT. On the other hand, in the large-deviation regime where $X$ scales linearly in $N$, i.e., $|X|\sim N$, we find that the final position is dominated by a single large displacement, a phenomenon also observed in the recent literature on anomalous transport \cite{embrechts_book,VBB19,MPG20,HB21}. Thus, in this region the PDF of $X$ can be written as
\begin{equation}
P(X,N)\simeq N \mP(X)\,,
\end{equation}
where $\mP(x)$ is the single-run PDF, given in Eq. \eqref{x_marg.2}, and the factor $N$ comes from the fact that any of the $N$ i.i.d. displacements can be the condensate. Thus, for $|X|\sim N$, $P(X,N)$ has a stretched exponential tail [see Eq. \eqref{eq:mP-asymptotic}]
\begin{equation}
P(X,N)\sim e^{-(3/2)|X|^{2/3}}\,.
\label{stretched_tail}
\end{equation}

To identify the correct scale at which the crossover between these two regimes occurs we match the Gaussian weight in Eq. \eqref{gaussian_tail} with the stretched-exponential tail in Eq. \eqref{stretched_tail}
\begin{equation}
e^{-X^2/(4N)}\sim e^{-(3/2)|X|^{2/3}}\,,
\end{equation}
yielding $|X|\sim N^{3/4}$. Thus, we zoom in this region $|X|\sim N^{3/4}$ and set $X=z N^{3/4}$, where $z$ describes the scaled position in the vicinity of the transition. In this intermediate regime, the distribution of $X$ assumes an anomalous large-deviation form. We observe that the PDF $P(X,N)$ is described by a Gaussian probability weight up to some critical value $X_c=z_c  N^{3/4} $ (where $z_c$ is a constant of order one), far outside of the region predicted by the CLT. At this critical point, the system undergoes a first-order condensation transition, signaled by a discontinuity in the first derivative of the rate function $F(z)$, where $z=X/N^{3/4}$.

\begin{figure}
\includegraphics[width=\columnwidth]{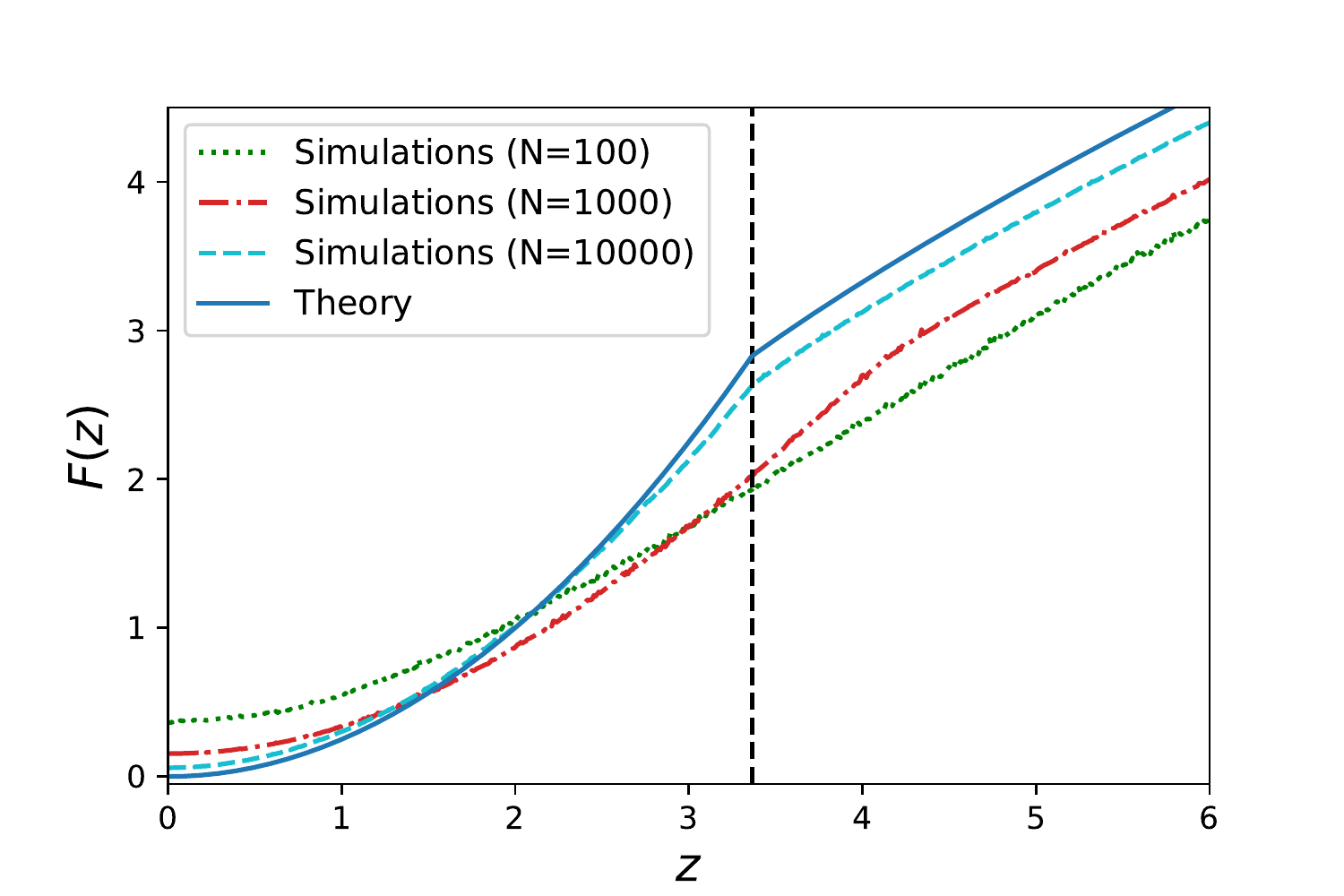}
\caption{The rate function $F(z)$ versus $z=X/N^{3/4}$. The continuous blue line represents the theoretical result in Eq. \eqref{F_intro}, while the dashed lines depict the results of numerical simulations, for different values of $N$. The first derivative of $F(z)$ is discontinuous at the critical point $z_c=2^{7/4}$, marked by a dashed vertical line.}
\label{fig:F}
\end{figure}

These different regimes can be summarized as follows
\be
P(X,N)\approx
\begin{cases}
\exp\left[-\frac{ X^2}{4  N}\right]& \text{for } |X|\sim \sqrt{N},\\
\\

\exp\left[-\sqrt{N}F \left(\frac{ |X|}{ N^{3/4}}\right) \right]& \text{for } |X|\sim N^{3/4},\\
\\
\exp\left[-\frac{3}{2} {|X|}^{2/3}\right]& \text{for } |X|\sim N,
\end{cases}
\label{regimes}
\ee
where 
\be
F(z)=
\begin{cases}
z^2/4 & \text{for } z<z_c,\\
\\
\chi(z) & \text{for } z>z_c,
\end{cases}
\label{F_intro}
\ee
with $z_c= 2^{7/4}$. Note that, for $z<z_c$, $P(X,N)$ is still described by the same Gaussian weight as in the typical regime. The function $\chi(z)$ can be computed only in the region $z>z_{\ell}=4 ~(2/3)^{3/4}$. Luckily, it turns out that $z_{\ell}=2.9511\ldots<z_c=3.3635\ldots$ and thus we find the exact expression of $\chi(z)$ in the region of interest $z>z_c$.
The full expression of $\chi(z)$ is rather complicated and is given in Eq. (\ref{chi_expression}) of Appendix \ref{app:chi}. Its asymptotic behavior is given by 
\be
\chi(z)=
\begin{cases}
\sqrt{6} +o(1)& \text{when } z\to z_{\ell}\,, \\
\\
\frac32 z^{2/3}- z^{-2/3}+o(z^{-2/3})& \text{when } z\to \infty.
\end{cases}
\label{asymptotics_chi}
\ee

From the second line of Eq. \eqref{asymptotics_chi} we observe that
the rate function can be approximated as $F(z)\simeq (3/2)z^{2/3}$ for
large $z$, smoothly connecting to the extreme large deviation regime
[see the third line of Eq. \eqref{regimes}]. From Eq. \eqref{F_intro},
we observe that the rate function $F(z)$ is singular at the critical
point $z=z_c$. Since the first derivative of $F(z)$ is discontinuous
at $z=z_c$, we say that the system undergoes a first-order phase
transition at $z=z_c$. The exact result for the rate function
$F(z)$ is shown in Fig.~\eqref{fig:F} (continuous blue line), and is in good agreement with numerical simulations (dotted lines), performed with a constrained Markov chain Monte Carlo (MCMC) algorithm (for the details of the
numerics see Appendix \ref{app:Monte-Carlo}).

\begin{figure}
\includegraphics[width=\columnwidth]{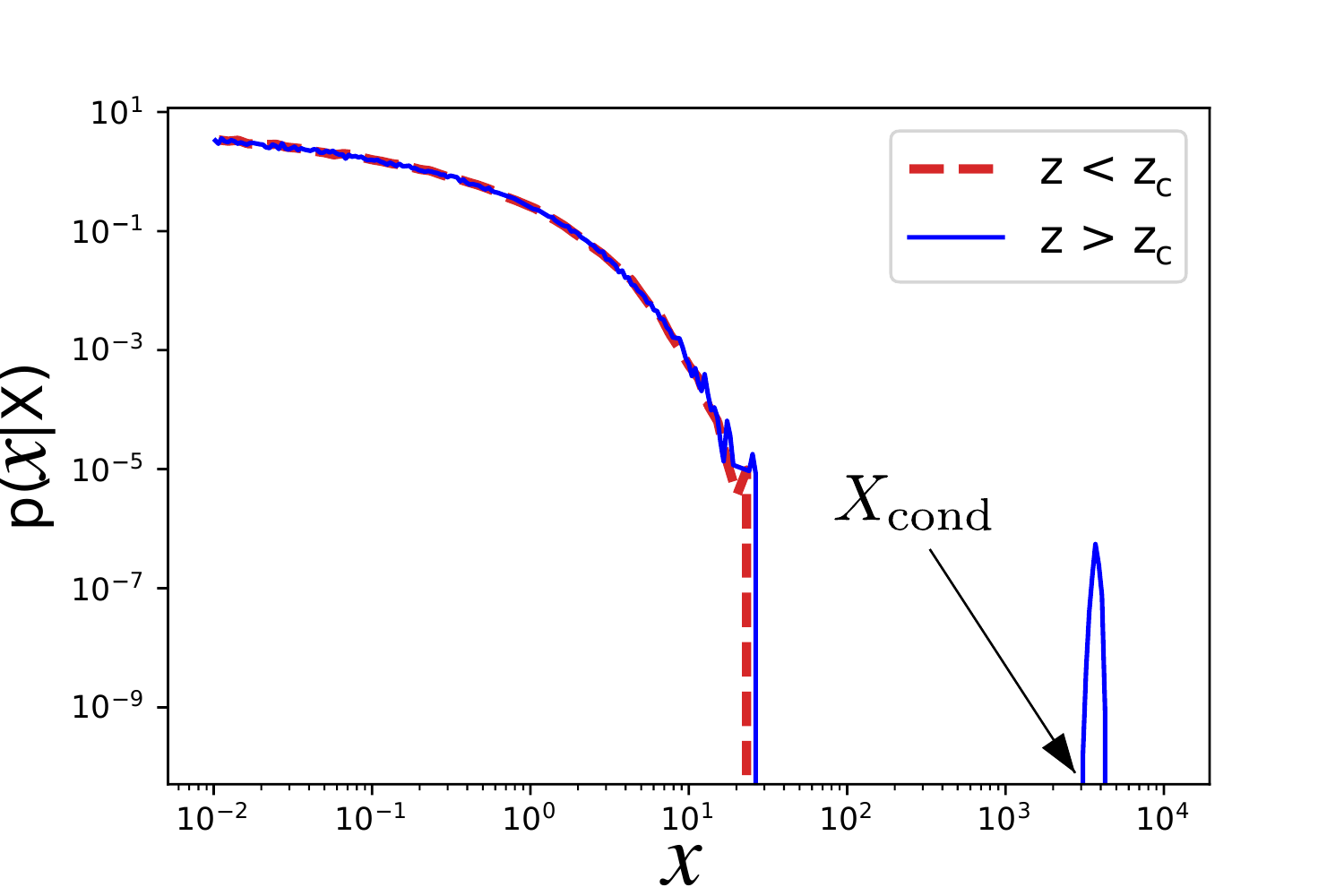}
\caption{Marginal probability $p(x|X)$ of a single-run displacement as a function of $x$, obtained from numerical simulations with $N=10^4$. The dashed red line corresponds to numerical simulations performed in the fluid phase, for $z<z_c$, where $z=X/N^{3/4}$. For $z>z_c$ (continuous blue line) the system is in the condensed phase and a bump appears in the tail of $p(x|X)$ at $x=X_{\rm cond}$, where $X_{\rm cond}=y^*(z)N^{3/4}$ and $y^*(z)$ is given in Eq. \eqref{y_expression}.}
\label{fig:marg}
\end{figure}

\subsubsection{Marginal probability distribution of a single-run displacement}

To gain insights into the nature of this first-order transition,
we present a detailed study of the marginal probability
distribution $p(x|X,N)$ of a single displacement conditioned on the
total displacement $X$ after $N$ steps. For the sake of simplicity we will use in the rest of the article the notation $p(x|X)$. However, it should be remembered that the marginal distribution depends also on the number of steps $N$. The variable $x$ can be any of the
displacements $x_1\,,\ldots\,,x_N$, since these variables are
i.i.d. Thus, given an RTP trajectory with total displacement $X$ and
choosing at random one of the $N$ runs, $p(x|X)$ describes the
distribution of the displacement $x$ associated to that run. For
simplicity we limit our discussion to the case $X>0$: all arguments hold identically for $X<0$, since $P(X,N)$ is symmetric around the origin.

The marginal probability $p(x|X)$, obtained from numerical simulations, is shown in Fig. \ref{fig:marg} as a function of $x$ for two different values of $z=X/N^{3/4}$. For $z<z_c$, we observe that $p(x|X)$ decays monotonically as a function of $x$. This observation is in agreement with the fact that in the fluid phase we expect each run to provide an order-one contribution to the total displacement. Upon crossing the critical point $z=z_c$, an additional bump appears in the tail of $p(x|X)$, signalling the presence of a condensate. The position of the bump scales with $N$ as $N^{3/4}$ and it has Gaussian fluctuations of order $\sqrt{N}$.

It turns out that for $x\sim O(1)$ the PDF $p(x|X)$ is given to leading order by
\be 
p(x|X)\simeq \mP (x)
\ee
where $\mP(x)$ is given in Eq. \eqref{x_marg.2}. In other words, when $x\sim O(1)$, the marginal distribution $p(x|X)$ is simply given by the unconstrained PDF $\mP(x)$. Note that this is valid for any $z>0$, i.e., both in the fluid and the condensed phases. Indeed, in Fig. \ref{fig:marg} we observe that the two numerical lines, obtained for $z<z_c$ and $z>z_c$, collapse into the same curve when $x\sim O(1)$.

\begin{figure}
\includegraphics[width=\columnwidth]{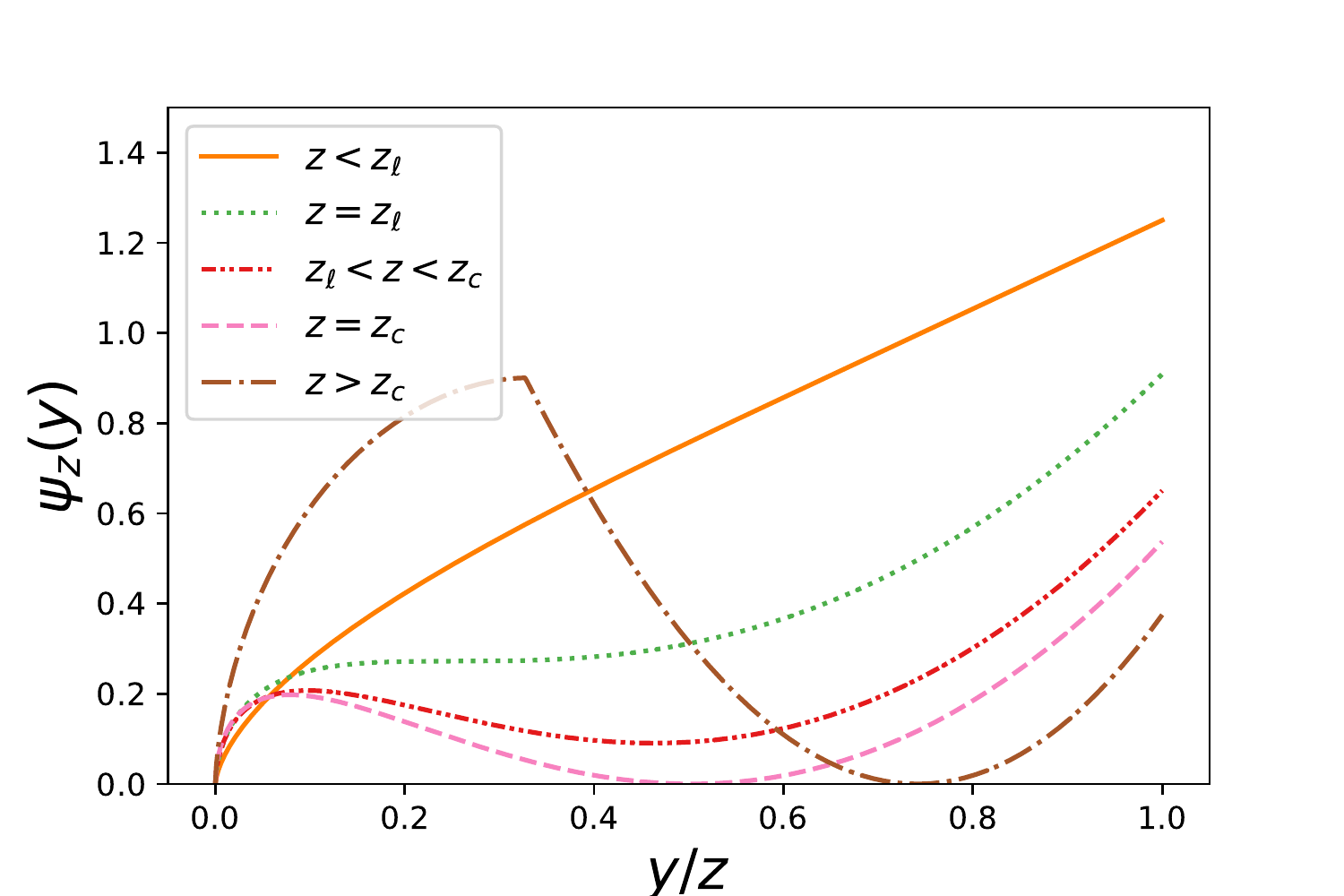}
\caption{The rate function $\psi_z(y)$, defined in Eq. \eqref{A_definition}, as a function of $y/z$, for different values of the $z$. For $z<z_{\ell}$, $\psi_z(y)$ has no minimum for $y>0$. At $z=z_{\ell}$ a minimum appears at $y^*>0$ with $\psi_z(y^*)>0$. Increasing $z$ further, the value of $\psi_z(y^*)$ decreases until for $z=z_c$ one finds that $\psi_z(y^*)=0$. For $z>z_c$, one still has that $\psi_z(y^*)=0$. Note that in this figure we plot $\psi_z(y)$ as a function of $y/z$ so that curves corresponding to different values of $z$ have the same range of values, since $0<y<z$.}
\label{fig:A}
\end{figure}

Thus, in order to distinguish between fluid and condensed phase one has to study the tail behavior of $p(x|X)$. In particular, in the region where $x\sim O(N^{3/4})$ we find that
\begin{equation}
p(x|X)\sim 
\exp\left[-\sqrt{N}\psi_z\left(\frac{x}{ N^{3/4}}\right)\right]\,,
\end{equation}
where $z=X/ N^{3/4}$,
\begin{equation}
\psi_z(y)=\frac32 y^{2/3}+F(z-y)-F(z)\,,
\end{equation}
and $F(z)$ is the rate function defined in Eq.~(\ref{F_intro}).
Thus, the probability of the rare fluctuations where $x\sim N^{3/4}$ is described by the rate function $\psi_z(y)$ (shown in Fig. \ref{fig:A}), where $y=x/N^{3/4}$.

For $z<z_c$, the function $\psi_z(y)$ is always positive and thus the probability of configurations with $x\sim N^{3/4}$ decays as $e^{-c\sqrt{N}}$, where $c>0$ is some positive constant. This can be observed in Fig. \ref{fig:marg}, where the empirical PDF $p(x|X)$ vanishes in the region $x\sim O(N^{3/4})$ for $z<z_c$. Conversely, for $z>z_c$ we find that there is a unique point $y^*>0$ at which $\psi_z(y^*)=0$. This zero of the rate function corresponds to the isolated bump in the tail of $p(x|X)$ as shown in Fig. \ref{fig:marg}. The bump is located at $x=X_{\rm cond}$, where
\be 
X_{\rm cond}=y^*(z) N^{3/4}\,,
\label{Xcond}
\ee 
where $y^*(z)$  is given in Eq. \eqref{y_expression}.

Indeed, by expanding $p(x|X)$ in the vicinity of $x=X_{\rm cond}$, we obtain
\begin{equation}
p(x|X)\simeq
p_{\rm cond}(x-X_{\rm cond},N)\,,
\end{equation}
where
\be
p_{\rm cond}(y,N) =\sqrt{\frac{\psi''_z(y^*)}{2\pi}} \frac{1}{N^{3/2}}\exp\left[-\frac{\psi''_z(y^*)  y^2}{2N}\right]\,,
\ee
and $\psi''_z(y)$ is the second derivative of $\psi_z(y)$ with respect to $y$. Thus, above the transition a bump appears in the tail of the marginal distribution at $x\simeq X_{\rm cond}$, corresponding to a condensate with Gaussian fluctuations of order $\sqrt{N}$ (see Fig. \ref{fig:marg}). Additionally, we show that
\be 
\int_{-\infty}^{\infty}dy\,p_{\rm cond}(y,N)=\frac{1}{N}\,,
\label{area}
\ee
in agreement with the fact that any of the i.i.d. variables $x_1\,,\ldots\,,x_N$ can become the condensate. In other words, in the condensed phase, $N-1$ displacements give an order-one contribution to the final position $X$, while a single displacement contributes to a finite fraction of $X$.

\begin{figure}
\includegraphics[width=\columnwidth]{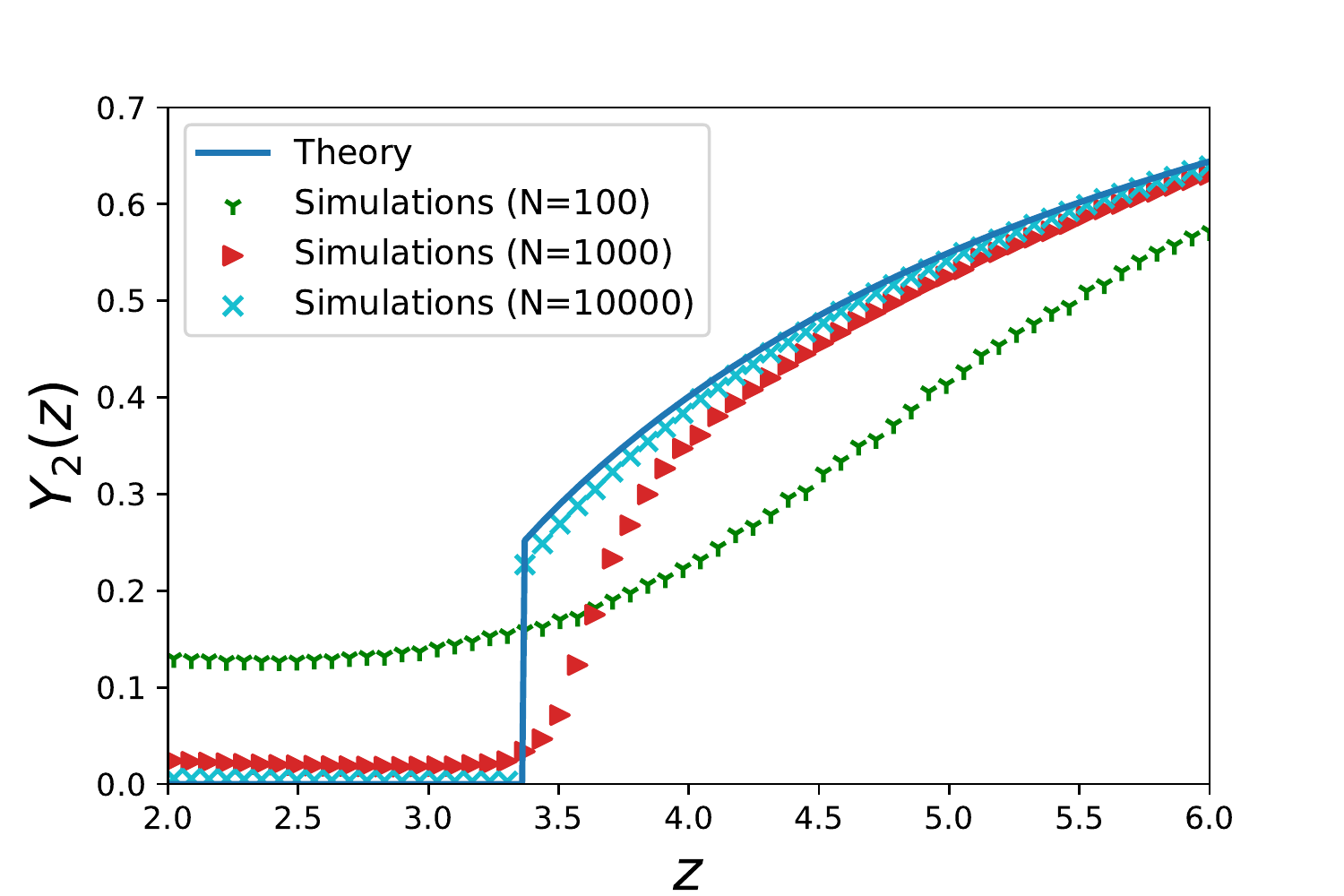}
\caption{The participation ratio $Y_2(z)$ as a function of the rescaled variable $z=X /N^{3/4}$. The continuous blue line represents the theoretical result in Eq. \eqref{Y2_expres}, while the symbols mark the results of numerical simulations, for different values of $N$. The participation ratio vanishes for $z<z_c=2^{7/4}$, while it is positive for $z>z_c$. At the critical value $z=z_c$, $Y_2(z)$ has a jump discontinuity.}
\label{fig:Y2}
\end{figure}

\subsubsection{Order parameter: the participation ratio}
Finally, we identify an order parameter for this first-order transition: the \emph{participation ratio} $Y_2(z)$, defined as
\be
Y_2(z)=\left\langle \frac{\sum_{i=1}^{N}x_i^2}{\left(\sum_{i=1}^{N}x_i\right)^2}\right\rangle_z\,,
\label{Y2}
\ee
where $\langle\ldots\rangle_z$ denotes the statistical average of the distribution of the displacements $x_1\,\ldots\,x_N$, conditioned on the event $X=z N^{3/4}$. For $z<z_c$, the variables $x_1\,,\ldots\,,x_N$ contribute to the total displacement $X$ by roughly the same amount of order one. Thus, the numerator in Eq. \eqref{Y2} scales as $O(N)$ while the denominator is equal to $X^2\sim O(N^{3/2})$ and hence $Y_2(z)\sim O(1/\sqrt{N})$.  On the other hand, in the condensed phase $z>z_c$ one single variable absorbs a finite fraction of $X$, while the other $N-1$ variables remain of order one. Hence, both the denominator and the numerator in Eq. \eqref{Y2} scale as $O(N^{3/2})$ and we then expect $Y_2(z)\sim O(1)$. Indeed, we show that, in the large-$N$ limit,
\begin{equation}
Y_2(z)=\begin{cases}
0 & \text{for } z<z_c,\\
\\
\left[y^*(z)/z\right]^2 & \text{for } z>z_c\,,
\end{cases}
\label{Y2_expres}
\end{equation}
where $y^*(z)$ is a given in Eq. \eqref{y_expression}. The expression for $Y_2(z)$ in Eq. \eqref{Y2_expres} is shown in Fig. \ref{fig:Y2} and is in good agreement with numerical simulations. The participation ratio $Y_2(z)$ is the natural order parameter of the system. Indeed, $Y_2(z)$ is zero below the transition while it becomes non-zero for $z>z_c$. Notably, $Y_2(z)$ has a jump discontinuity at the critical value $z=z_c$, implying a first-order transition.

We show that the $Y_2(z)$ is related to the condensate fraction $m_c$, i.e., the fraction of the total displacement $X$ which is carried by the condensate, by the simple relation
\begin{equation}
Y_2(z)=m_c^2\,.
\end{equation} 
We also compute the asymptotic behavior of $Y_2(z)$ in the region $z>z_c$, showing that
\begin{equation}
Y_2(z)\simeq\begin{cases}
1/4+2^{-7/4}(z-z_c) & \text{for } z\to z_c^+,\\
\\
1-4 z^{-4/3} & \text{for } z\to\infty.
\end{cases}
\label{mc_lim}
\end{equation}
We observe that, above the transition, a condensate forms and the participation ratio jumps to the finite value $1/4$, corresponding to $m_c=1/2$, i.e., to a configuration where the condensate absorbs half of the total displacement $X$. Increasing $z$ further, the participation ratio $Y_2(z)$ increases and it goes to 1 as $z\to\infty$. Note that a participation ratio equal to 1 corresponds to a configuration where the whole displacement $X$ is absorbed by a single jump.

\begin{figure}
%%%
%%%
\centering 

\includegraphics[scale=1]{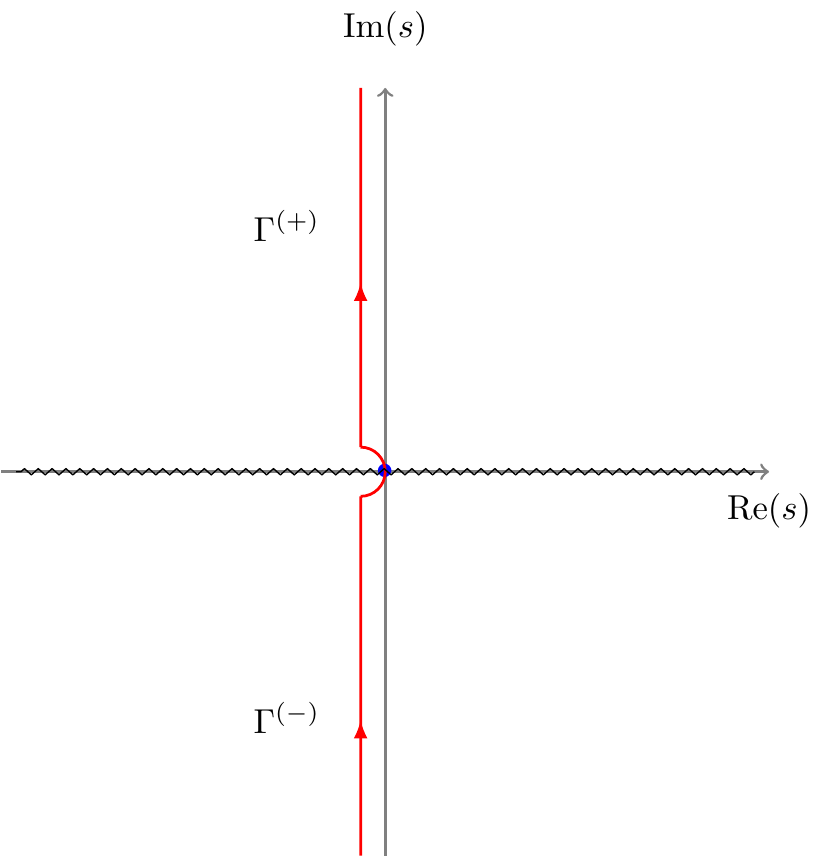} 
\caption{Analytic properties in the complex-$s$ plane of the function $\mL(s)$, given in Eq.~(\ref{L_tilde}). The function $\mL(s)$ is characteirzed by two branch cuts (black wiggled lines), which correspond to the intervals $]-\infty,0[$ and $]0,\infty[$ in the real axis. The full red line represents the contour $\Gamma$ that we choose to perform the integral in Eq. \eqref{eq:inv-Lapl-transf2} (in the case $X>0$).}
\label{fig:Bromwich}
\end{figure}

\section{Position distribution}

\label{sec:position}

In this Section, we investigate the PDF $P(X,N)$ of the position $X$ of the RTP after $N$ running phases. By using the integral representation
of the delta function
\be
\delta(X)= \frac{1}{2\pi i}\int_{\Gamma} e^{- s\, X}\, ds\,,
\ee
where $\Gamma$ is the imaginary-axis Bromwich contour in the complex $s$ plane (see Fig. \ref{fig:Bromwich}), one can decouple the integrals over the variables $x_i$  in Eq. (\ref{eq:PXN-free}) and rewrite $P(X,N)$ as
\be
P(X,N) = \frac{1}{2\pi i} \int_{\Gamma} ds~e^{sX}\left[\mL(s)\right]^N\,.
\label{eq:inv-Lapl-transf2a}
\ee
where
\be
\mL(s)= \int_{-\infty}^{\infty} dx\, e^{-s\, x}\, \mP(x)\, .
\label{LT.1}
\ee
The expression in Eq. \eqref{eq:inv-Lapl-transf2a} can be then rewritten as
\be
P(X,N) = \frac{1}{2\pi i} \int_{\Gamma} ds~e^{sX+N\log[\mL(s)]}\,.
\label{eq:inv-Lapl-transf2}
\ee

The function $\mL(s)$ can be explicitly computed by substituting the expression for $\mP(x)$, given in Eq. \eqref{x_marg.2}, into Eq. \eqref{LT.1}, yielding
\be
\mL(s)= \sqrt{\pi} \frac{e^{-1/(2s^2)}}{\sqrt{-2 s^2}}~\textrm{erfc}\left[\frac{1}{\sqrt{-2s^2}}\right].
\label{L_tilde}
\ee
From a direct inspection of Eq. \eqref{L_tilde}, one finds that the contour integral in Eq. \eqref{eq:inv-Lapl-transf2} cannot be computed directly using a saddle point approximation. Indeed, looking at the analytic structure of the function $\mL(s)$ in the complex plane in
Fig.~\ref{fig:Bromwich}, we observe that $\mL(s)$ is non-analytic on the full real line in the complex-$s$ plane, except at $s=0$ where it has a removable discontinuity with $\mL(0)=1$. In particular, the function $\mL(s)$ is characterized by two branch cuts (the black wiggled lines in Fig. \ref{fig:Bromwich}) in the intervals $]-\infty,0[$ and $]0,\infty[$ in the real-$s$ axis. Our main results for $P(X,N)$ are obtained by analyzing the contour integral in Eq. \eqref{eq:inv-Lapl-transf2} in the large-$N$ limit.

\subsubsection{Typical regime}

Let us first consider the typical regime where $X= z \sqrt{N}$, where the rescaled variable $z$ can be positive or negative. To investigate this regime we can choose the contour of integration $\Gamma$ to lie on the imaginary axis in the complex $s$ plane. Performing the change of variable $s\to s/\sqrt{N}$ in Eq. \eqref{eq:inv-Lapl-transf2}, we obtain
\be
P(X=z \sqrt{N},N) = \frac{1}{2\pi i} \frac{1}{\sqrt{N}}
\int_{\Gamma} ds~e^{sz+N\log[\mL(s/\sqrt{N})]}.
\label{eq:regime1}
\ee
Expanding the expression of $\mL(s)$, given in Eq. \eqref{L_tilde}, for small $|s|$ with $\operatorname{Re}(s)=0$, we find
\begin{equation}
\mL(s)=1+s^2+o(s^2)\,.
\label{eq:expansion1}
\end{equation}
Using this expansion in Eq. \eqref{eq:regime1}, we find that for large $N$
\be
P(X= z \sqrt{N},N) \simeq\frac{1}{2\pi i} \frac{1}{\sqrt{N}}
\int_{\Gamma} ds~e^{sz+s^2}.
\label{eq:regime1_2}
\ee
Performing the Gaussian integral, we finally get for $N\gg 1$ and $|X|\sim \sqrt{N}$
\be
P(X,N) \simeq \frac{1}{2\sqrt{\pi N}}\exp\left(- X^2/4  N\right)\,,
\label{eq:regime1_3}
\ee
as given in the first line of Eq. (\ref{regimes}). Note that this result in Eq. \eqref{eq:regime1_3} is a simple consequence of the CLT.

\subsubsection{Extreme large deviations}

We now consider the regime of extreme large deviations, where $X$ scales linearly with $N$. In the following, we will focus on the case $X>0$. The complementary case $X<0$ can be obtained by the symmetry $P(X,N)=P(-X,N)$.

To extract the atypical behavior of the PDF $P(X,N)$ in this regime from the integral representation in Eq. \eqref{eq:inv-Lapl-transf2}, it is useful to deform the contour $\Gamma$, as shown in Fig. \ref{fig:Bromwich}. To be precise, we choose the contour to run  parallel to the imaginary axis with $\operatorname{Re}(s)<0$. Note that the contour needs to be deformed to pass through the origin, due to the presence of the two branch cuts (see Fig. \ref{fig:Bromwich}). 

It is useful to define the variable $z= X/ N>0$. Performing the change of variable $s\to s/N$ in Eq. \eqref{eq:inv-Lapl-transf2}, we find
\be
P(X= z N,N) = \frac{1}{2\pi i} \frac{1}{N}
\int_{\Gamma} ds~e^{sz+N\log[\mL(s/N)]}.
\label{eq:regime2}
\ee
We now need to expand $\mL(s)$ for small $|s|$. We find that, when $\operatorname{Im}(s)<0$ and $\operatorname{Re}(s)<0$,
\be
\mL(s)\simeq 1+s^2+o(s^2)\,.
\label{expans1}
\ee
On the other hand, when $\operatorname{Im}(s)>0$ and $\operatorname{Re}(s)<0$, we obtain
\be
\mL(s)\simeq 1+s^2+\frac{\sqrt{2\pi}}{\sqrt{-s^2}}e^{-1/(2s^2)}+ o(s^2)\,.
\label{expans2}
\ee
When $\operatorname{Im}(s)>0$, one has an additional non-analytic term in the expansion of $\mL(s)$. Thus, it is useful to write the contour $\Gamma$ as the union of the contours $\Gamma^+$, in the positive imaginary semiplane, and $\Gamma^-$, in the negative imaginary semiplane. Plugging the expansions of $\mL(s)$, given in Eqs. \eqref{expans1} and \eqref{expans2}, into Eq. \eqref{eq:regime2}, we find
\bea
&&P(X= z N,N) \simeq \frac{1}{2\pi i} \frac{1}{N}
\int_{\Gamma^-} ds~e^{sz+s^2/N}\\
&+& \frac{1}{2\pi i} 
\frac{1}{N}\int_{\Gamma^+} ds~\exp\left[sz+\frac{s^2}{N} +N^2\frac{\sqrt{2\pi}}{\sqrt{-s^2}}e^{-N^2/(2 s^2)}\right]\,.\nonumber
\label{eq:regime2_2}
\eea
Expanding for large $N$, we obtain
\bea
&& P(X= z N,N) \simeq \frac{1}{2\pi i}\frac{1}{N}
\int_{\Gamma^-} ds~e^{sz}\left(1+\frac{s^2}{N}\right)\\
&+& \frac{1}{2\pi i} \frac{1}{N}
\int_{\Gamma^+} ds~e^{sz}\left[1+\frac{s^2}{N}+N^2\frac{\sqrt{2\pi}}{\sqrt{-s^2}}e^{-N^2/(2 s^2)}\right]\,.\nonumber
\label{eq:regime2_3}
\eea
Regrouping the different terms we obtain
\be
\begin{split}
P(X= z N,N) \simeq \frac{1}{2\pi i} \frac{1}{N}
\int_{\Gamma} ds~e^{sz}\left(1+\frac{s^2}{N}\right)\\
+ \frac{N}{\sqrt{2\pi} i} \int_{\Gamma^+} ds~\exp\left[ sz-N^2/(2 s^2)\right] \frac{1}{\sqrt{-s^2}}\,.
\end{split}
\ee
It is possible to show that the term
\be
\frac{1}{2\pi i} \frac{1}{N}
\int_{\Gamma} ds~e^{sz}\left(1+\frac{s^2}{N}\right)
\ee
vanishes for any $z\neq 0$. Thus, we are left with the integral over the contour $\Gamma^+$. To perform this integral, we rotate the contour anticlockwise by an angle of $\pi/2$ and we obtain, using the parametrization $s=ik$
\be
P(X=z N,N) \simeq
 \frac{N}{\sqrt{2\pi} } \int_{-\infty}^{0} \frac{dk}{k}~\exp\left[ i k z+N^2/(2 k^2)\right]\,.
\label{eq:regime2_4}
\ee
This integral can be computed exactly and we obtain in the regime where $X\sim N$
\be
P(X,N)\simeq N \frac{1}{2\sqrt{2}\pi} G_{0,3}^{3,0} \left(\begin{array}{c|ccc} & 0 & 0 & 0 \\ \frac{ X^2}{8 }&  &  &  \\  & 0 & 0 & 1/2 \end{array}\right),
\ee
where $G_{0,3}^{3,0}(x|\ldots)$ is the Meijer G-function.
Comparing this expression to the one for the marginal PDF $\mP(x)$ of the single displacements [see Eq. \eqref{x_marg.2}], we find that
\be
P(X,N)\simeq N \mP(X)\,.
\label{eq:regime2_5}
\ee
Finally, using the large-$x$ expansion of $\mP(x)$, given in Eq. \eqref{eq:mP-asymptotic}, we obtain
\be
P(X,N)\simeq N  \frac{1 }{\sqrt{3}\,
  |X|^{1/3}}\, e^{-\frac{3}{2}\, {|X|}^{2/3}}\,,
\label{eq:regime2_6}
\ee
as we anticipated in Eq. \eqref{regimes}.

The result in Eq. \eqref{eq:regime2_5} can be interpreted as follows. In the extreme large deviation regime, the final position $X$ is dominated by a single large displacement, which has probability weight $\mP(X)$. Since this atypical displacement can be any of the $N$ variables $x_1\,,\ldots\,,x_N$, the factor $N$ is also present.

\subsubsection{Anomalous large deviations: matching regime}

Finally, let us consider the intermediate regime where $X\sim N^{3/4}$, which interpolates between the typical regime and the extreme large-deviation regime. As discussed in Sec. \ref{sec:model-results}, this unusual scale $N^{3/4}$ can be obtained by matching the exponent $X^2/N$ of the expression of $P(X,N)$ in the typical regime and the exponent $X^{2/3}$ of the extreme large-deviation regime [see the third line in Eq. \eqref{regimes}].

We will limit our discussion to the case $X>0$. We consider
  again the contour $\Gamma$ shown in Fig. \ref{fig:Bromwich}. We
  define the variable $z= X/ N^{3/4}>0$, so that what we have to
  compute is:
  \begin{align}
P(X= z N^{3/4},N) = \frac{1}{2\pi i}  \int_{\Gamma} ds~e^{szN^{3/4}+N\log[\mL(s)]}.
\end{align}
By expanding $\log[\mL(s)]$ around the origin, using the expressions in Eqs. (\ref{expans1}) and (\ref{expans2}), we get
  \begin{align}
  & P(X= z N^{3/4},N) \simeq \frac{1}{2\pi i} \int_{\Gamma^-} ds~e^{szN^{3/4} +N s^2} + \nonumber \\
  & + \frac{1}{2\pi i} \int_{\Gamma^+} ds~e^{sz N^{3/4}+N s^2+N \sqrt{2\pi} e^{-1/(2 s^2)}/\sqrt{-s^2}}\,.
    \label{eq:PX_split_1}
  \end{align}
We then expand 
  \begin{align}
 \exp\left[N \sqrt{2\pi}~\frac{e^{-1/(2 s^2)}}{\sqrt{-s^2}} \right]~\simeq~
    1 + N \sqrt{2\pi}~\frac{e^{-1/(2 s^2)}}{\sqrt{-s^2}} \,,
  \end{align}
so that in we can rewrite $P(X,N)$ as the sum of a Gaussian term and an anomalous term
\be
P(X,N) \simeq P_G(X,N) +P_A(X,N)\,.
\label{eq:regime3_3}
\ee
Where the Gaussian term is given by
\be
P_G(X,N)=\frac{1}{2\pi i} \int_{-i\infty}^{i\infty} ds~e^{sz N^{3/4}+N s^2}
\label{gaussian_term}
\ee
and the anomalous term reads
\bea
P_A(X,N)
=\frac{N}{i} \int_{\Gamma^+} ds~\frac{1}{\sqrt{-s^2}}~e^{szN^{3/4}+N s^2-1/(2s^2)}.\nonumber \\ 
\label{anomalous_term}
\eea 

Performing the Gaussian integral in Eq. \eqref{gaussian_term}, we find 
\begin{align}
  P_G(X,N)= \frac{1}{2\sqrt{\pi N}}~e^{-\sqrt{N} z^2/4}\,.
  \label{gaussian_term_2}
\end{align}
In order to evaluate the integral in Eq. \eqref{anomalous_term} we first perform the change of variable $s\to s/N^{1/4}$, yielding
\begin{align}
P_A(X,N)
=\frac{N}{i} \int_{\Gamma^+} ds~\frac{1}{\sqrt{-s^2}} e^{\sqrt{N}G_z(s)}\,,
\label{anomalous_term2}
\end{align}
where
\be G_z(s)=z s+s^2-\frac{1}{2s^2}\,.
\ee
It turns out that the integral in Eq.~\eqref{anomalous_term2} can be computed by means of saddle-point approximation only for $z>z_{\ell}=4(2/3)^{3/4}$
(see Appendix \ref{app:chi}). Indeed, the saddle point equation
\be
G'_z(s)=z+2s+\frac{1}{s^3}=0
\ee
has real solutions in $s$ only for $z>z_{\ell}$.  Thus, for $z>z_{\ell}$ it is possible compute the
integral in Eq. \eqref{anomalous_term2} by saddle point method and we
obtain \be P_A(X=z N^{3/4},N)\sim e^{-\sqrt{N}\chi(z)}\,,
\label{anomalous_term_3}
\ee
where the function $\chi(z)$ is computed exactly in Appendix \ref{app:chi} for $z>z_{\ell}$ and is given in Eq. (\ref{chi_expression}). Note that for $z<z_{\ell}$ the integral in Eq. \eqref{anomalous_term2}, even if hard to evaluate, is still well defined.

Plugging the expressions for $P_A(X,N)$ and $P_G(X,N)$, given in Eqs. (\ref{gaussian_term_2}) and (\ref{anomalous_term_3}), into Eq. (\ref{eq:regime3_3}), we find that
\be
P(X= z N^{3/4},N)\sim e^{-\sqrt{N} z^2/4}+ e^{-\sqrt{N}\chi(z)}\,.
\label{sum_anom_gauss}
\ee
From this expression it is clear that for large $N$ the two terms will compete, since the two exponents both scale as $\sqrt{N}$. In particular, for large $N$ we find that
\be
P(X= z N^{3/4},N)\sim e^{-\sqrt{N} F(z)}\,,
\ee
where 
\be 
F(z)=\min\left[\frac{z^2}{4},\chi(z)\right]\,,
\label{F}
\ee
where we know the expression of $\chi(z)$ only for $z>z_{\ell}$.  Luckily, it turns out that $\chi(z)<z^2/4$ only for $z>z_c=2^{7/4}$ and that $z_c>z_{\ell}$ (see Appendix \ref{app:zc}). Thus, we know the exact expression of $\chi(z)$ in the relevant region $z>z_{c}$ and we obtain
\be
F(z)=
\begin{cases}
z^2/4 & \text{for } z<z_c,\\
\\
\chi(z) & \text{for } z>z_c.
\end{cases}
\label{F_cases}
\ee
This rate function $F(z)$ is shown in Fig. \ref{fig:F} and is in good agreement with numerical simulations. From the first line in Eq. \eqref{F_cases} it is clear that the probability $P(X,N)$ remains Gaussian outside of the typical regime and up to $X= z_c N^{3/4}$. Moreover, it is easy to check that $F'(z)$, i.e., the first derivative of the rate function, is discontinuous at $z=z_c$, corresponding to a first-order dynamical phase transition.

The expression in Eq. \eqref{sum_anom_gauss} clarifies the mechanism of the transition, which resembles a first-order phase transitions of classical thermodynamics. Indeed, the transition is the result of the competition between two phases: the fluid phase, whose probability is described by the Gaussian weight, and the condensed phase, associated with the anomalous weight. In particular, to each phase corresponds a rate function ($z^2/4$ for the fluid phase and $\chi(z)$ for the condensed phase), which plays the role of the free energy for out-of-equilibrium systems. At a given value of the control parameter $z$, the system will be in the phase with lower rate function. Thus, the critical point $z_c$ is by definition the value for which the two rate functions are equal.

It is also possible to compute the asymptotics of $\chi(z)$ at the limits of its domain $[z_{\ell},\infty[$ (see Appendix \ref{app:asymptotics}). For $z\to z_{\ell}$ one obtains
\be
\chi(z)=\sqrt{6} +o(1)\,,
\ee
while for $z\to \infty$ 
\be
\chi(z)=\frac{3}{2}z^{2/3}- z^{-2/3}+o(z^{-2/3})\,.
\label{asympt_2}
\ee
Using the expansion in Eq. \eqref{asympt_2} we find that, starting from the intermediate regime $X=z N^{3/4} $ and taking the limit $z\to \infty$, one get, to leading order, 
\be
P(X= z N^{3/4},N)\sim e^{-\sqrt{N}(3/2)z^{2/3}}\,.
\ee
Finally, using $z=X N^{-3/4}$, we obtain
\be
P(X,N)\sim e^{-(3/2) X^{3/2}}\,,
\ee
smoothly matching with the the expression of $P(X,N)$ in the extreme large deviation regime, where $X\sim N$ [see Eq. \eqref{regimes}].

\section{Marginal probability density and the participation ratio}
\label{sec:marginal}

In this section we consider the marginal distribution $p(x|X)$ of a single displacement $x$, conditioned on the value of the final position $X$. Note that $x$ could be any of the i.i.d. displacements $x_1\,,\ldots\,,x_N$, for example we can choose $x=x_1$. We focus on the intermediate regime $X=z N^{3/4}$ and $X>0$. In the subcritical fluid phase $z<z_c$ we expect this distribution to be peaked around order-one values of $x$, since the different displacements $x_1\,,\ldots\,, x_N$ contribute to the final position by roughly the same amount. On the other hand, in the condensed phase $z>z_c$, we expect that one single displacement, which we will refer to as the \emph{condensate}, contributes extensively to the final position $X$. We denote by $m_c$ the fraction of the total displacement $X$ which is in the condensate. For $z>z_c$, we expect that a bump, corresponding to the condensate, develops in the tail of the marginal distribution $p(x|X)$. Since the condensate could be any of the $N$ displacements $x_1\,,\ldots\,, x_N$, we expect the area under this bump to be $1/N$.

Our starting point is the joint probability of the displacements $\{x_i\}=x_1\,,\ldots\,,x_N$ and of the final position $X=\sum_{i=1}^{N}x_i$, which is given by
\be 
p(\{x_i\},X)=\prod_{i=1}^{N}\mP(x_i)~\delta\left(X-\sum_{i=1}^Nx_i\right)\,,
\ee
where $\mP(x)$ is the unconstrained marginal probability, given in Eq. \eqref{x_marg.2}. From this expression, integrating over $x_2\,,\ldots\,,x_N$, we obtain the PDF of $x_1$ and $X$
\bea
p(x_1,X)&=&\mP(x_1)\int_{-\infty}^{\infty}dx_2\,\ldots\int_{-\infty}^{\infty}dx_N\left[\prod_{i=2}^{N}\mP(x_i)\right]\nonumber\\
&\times &\delta\left(X-x_1-\sum_{i=2}^N x_i\right)\,.
\label{joint_x_X}
\eea
Note that in principle one could choose, instead of $x_1$, any of the variables $x_1\,,\ldots\,,x_N$. Thus, from now on we will use the notation $x=x_1$. We notice that the term
\be
\int_{-\infty}^{\infty}dx_2\,\ldots\int_{-\infty}^{\infty}dx_N\left[\prod_{i=2}^{N}\mP(x_i)\right]\delta\left(X-x-\sum_{i=2}^N x_i\right)\,,
\ee
is exactly equal to $P(X-x,N-1)$, defined in Eq. \eqref{eq:PXN-free}. Thus, Eq. \eqref{joint_x_X} becomes
\bea
p(x,X)&=&\mP(x)P(X-x,N-1)\,.
\label{joint_x_X2}
\eea
Finally, since $p(x|X)$ is defined as the PDF of the single displacement $x$, conditioned on the final position $X$, we find 
\be
p(x|X)=\mP(x)\frac{P(X-x,N-1)}{P(X,N)}\,.
\label{marg_x|X}
\ee
Note that since the $N$ runs are identically distributed and the total displacement $X=\sum_{i}x_i$ is fixed, the first moment of $p(x|X)$ is given by
\begin{equation}
\int_{-\infty}^{\infty}dx~x~p(x|X)=\frac{X}{N}\,.
\label{first_moment}
\end{equation}
In particular, it is relevant to observe that even in the large-deviation regime, where $X\sim O(N^{3/4})$, the first moment of $p(x|X)$ vanishes as $N^{-1/4}$ for large $N$.

The main goal of this section is to analyze this marginal probability $p(x|X)$ in the intermediate large-deviation regime, where the dynamical phase transition occurs. In the previous section we have shown that, when $X=z N^{3/4}$, the distribution of $X$ is given, in the large-$N$ limit, by
\begin{equation}
P(X=z N^{3/4},N)\sim e^{-\sqrt{N}F(z)}\,,
\end{equation}
where $F(z)$ is given in Eq. \eqref{F_cases}. Plugging this expression into Eq. \eqref{marg_x|X}, for large $N$, we obtain 
\bea
p(x|X)\sim \mP(x)e^{-\sqrt{N}(F(z-y)-F(z))}\,,
\label{marg_x|X2}
\eea
where $y=x/ N^{3/4}$ and $z=X/N^{3/4}$. Thus, when $x\ll N^{3/4}$ the rescaled variable $y$ goes to zero for $N\to \infty$ and, expanding $F(z-y)$ for small $y$, we obtain
\bea
p(x|X)\simeq \mP(x) e^{x F'(z) N^{-1/4}}\,.
\label{marg_x|X3}
\eea
Thus, for $x\ll N^{3/4}$, we find that to leading order the marginal PDF $p(x|X)$ is simply given by the unconstrained PDF $\mP(x)$. The exponential correction factor in Eq. \eqref{marg_x|X3} skews the distribution $p(x|X)$ to the right (since $F'(z)>0$ for $z>0$). Thus, the value $F'(z)$ quantifies the asymmetry of $p(x|X)$.

\begin{figure}
\includegraphics[width=\columnwidth]{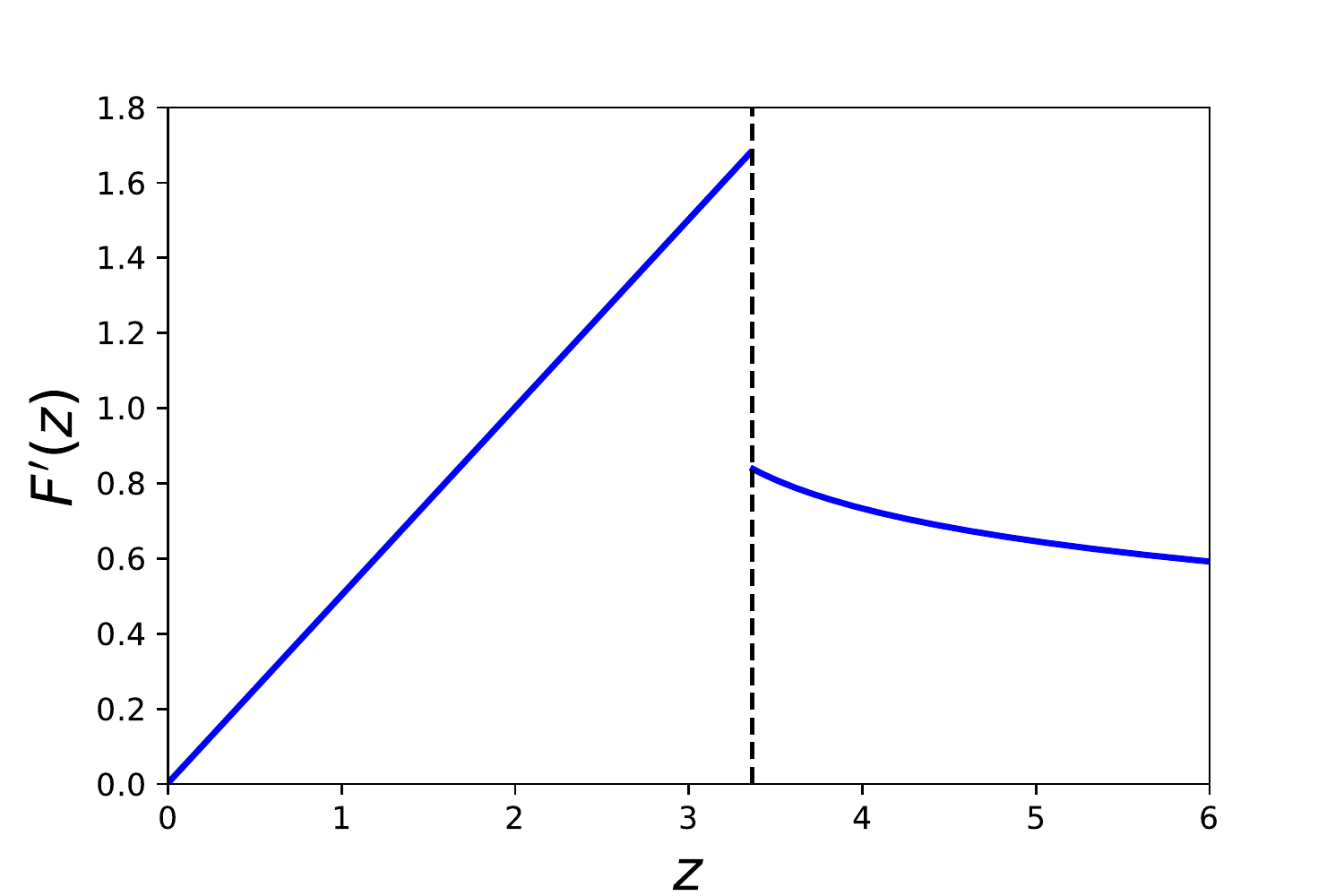}
\caption{First derivative of the rate function $F(z)$, given in Eq. \eqref{F}, versus $z=X/N^{3/4}$.}
\label{fig:Fprime}
\end{figure}

Indeed, computing the average value of the PDF in Eq. \eqref{marg_x|X3} and expanding for large $N$, we find
\begin{eqnarray}
&& \int_{-\infty}^{\infty}dx~x~\mP(x) e^{x F'(z) N^{-1/4}}\\
& \simeq & \int_{-\infty}^{\infty}dx~x~\mP(x) \left(1+xF'(z)N^{-1/4}\right)\,.\nonumber
\end{eqnarray}
Using the values of the first two moments of $\mP(x)$, given in Eqs. \eqref{mean_x} and \eqref{var_x}, we obtain
\begin{equation}
 \int_{-\infty}^{\infty}dx~x~\mP(x) e^{x F'(z) N^{-1/4}}\simeq m_f\frac{X}{N}\,,
 \label{avg_bulk}
\end{equation}
where we have defined 
\begin{equation}
m_f=\frac{2F'(z)}{z}\,.
\label{mf}
\end{equation}
We will call this quantity $m_f$ \emph{fluid fraction}, for reasons that will be clarified later in this section. Notably, using the expression of $F(z)$ in Eq. \eqref{F}, we obtain $m_f=1$ for $z<z_c$  and thus
\begin{equation}
 \int_{-\infty}^{\infty}dx~x~\mP(x) e^{x F'(z) N^{-1/4}}\simeq\frac{X}{N}\,,
\end{equation}
in agreement with the strict conservation law in Eq. \eqref{first_moment}. However, at $z=z_c$ the function $F'(z)$ is discontinuous and it decreases with $z$ for $z>z_c$ [see Fig. \eqref{fig:Fprime}]. Thus, for $z>z_c$, $m_f$ becomes a decreasing function of $z$, meaning that increasing the value of the total displacement $X$ the value of the typical single-run displacement $x$ decreases.

This apparent contradiction is a consequence of the fact that the expression in Eq. \eqref{marg_x|X3} is only valid for $x\ll N^{3/4}$ and that the tail of the distribution $p(x|X)$ could in principle also contribute to the mean value. Thus, it is useful to consider the regime $x\sim N^{3/4}$ and to define the scaled variable $y=x/ N^{3/4}\sim O(1)$. In this regime, plugging the large-$x$ expansion of $\mP(x)$, given in Eq. \eqref{eq:mP-asymptotic}, into Eq. \eqref{marg_x|X2}, we obtain
\be
p(x|X) \sim e^{-\sqrt{N}\psi_z(x/N^{3/4})}\,,
\label{marg_x|X4}
\ee
where
\begin{equation}
\psi_z(y)=\frac{3}{2}y^{2/3}+F(z-y)-F(z)\,,
\label{A_definition}
\end{equation}
and $F(z)$ is given in Eq. \eqref{F_cases}. We recall that $y=x/ N^{3/4}$ and $z=X /N^{3/4}$. This rate function $\psi_z(y)$, parametrized by $z$, describes the probability of the large deviations where $x\sim O(N^{3/4})$. One can check that $\psi_z(y)\geq 0$ for any $z>0$ and $y>0$ (see Fig. \ref{fig:A}).

It is instructive to investigate the behavior of $\psi_z(y)$ as a function of $y$, fixing $z$. For $z<z_{\ell}$, the function $\psi_z(y)$ is monotonic in $y$ (see Appendix \ref{app:location}). On the other hand, for $z>z_{\ell}$ a local minimum appears at some value $y=y^*>0$ [the precise value of $y^*$ depends on $z$ and is given in \eqref{y_expression}]. However, the value $\psi_z(y^*)$ corresponding to this minimum is initially strictly positive (see Fig. \ref{fig:A}). Thus, configurations where $x=y^* N^{3/4}$ are still exponentially rare for large $N$ and do not contribute to the average value of $x$.

Increasing $z$, this minimum $\psi_z(y^*)$ decreases, until, at the critical value $z=z_c$ it becomes zero. Thus, the configuration where $y=y^*$ becomes typical and a condensate develops at $x=X_{\rm cond}$, where $X_{\rm cond}=y^* N^{3/4}$. For $z>z_c$, there is always a unique value $y^*>0$ at which $\psi_z(y^*)=0$. This zero of the rate function $\psi_z(y)$ corresponds to the appearance of a bump in the tail of $p(x|X)$ and the value $y^*$ increases as a function of $z$. Thus, for $z>z_c$ the fraction $m_c=X_{\rm cond}/X$ of the total displacement $X$ that belongs to the condensate is given by 
\begin{equation}
m_c=\frac{y^*N^{3/4}}{X}=\frac{y^*}{z}\,.
\end{equation}
On the other hand, below the transition no condensate is present and thus $m_c=0$. To summarize, we obtain
\begin{equation}
m_c=\begin{cases}
0 & \text{for } z<z_c,\\
\\
y^*/z & \text{for } z>z_c\,,
\end{cases}
\label{mc_expression}
\end{equation}
where $y^*$ is given in Eq. \eqref{y_expression}. The condensate fraction $m_c$ is shown in Fig. \ref{fig:mc} as a function of $z$. In particular, in the region $z>z_c$ we obtain (see Appendix \ref{app:location})
\begin{equation}
m_c\simeq \begin{cases}
1/2 +(z-z_c)/2^{7/4}& \text{for } z\to {z_c}^+,\\
\\
1-2 z^{-4/3}& \text{for } z\to \infty~.
\end{cases}
\label{mc_limits}
\end{equation}
Thus, when crossing the transition line $z=z_c$, the fraction $m_c$ jumps from zero to $1/2$, signaling a first-order phase transition and a condensate, containing half of the total displacement $X$, appears. Increasing $z$ further, the fraction $m_c$ increases and it goes to one when $z\to\infty$. Thus, for $z\gg z_c$ almost the totality of the displacement $X$ is in the condensate.

The behavior of $\psi_z(y)$ clarifies the mechanism of the dynamical transition, which turns out to be reminiscent of equilibrium first-order transitions. Indeed, the rate function $\psi_z(y)$ can be interpreted as the free energy of the system, while the variables $y$ and $z$ are the order and the control parameters, respectively. The global minimum of $\psi_z(y)$ corresponds to the most probable value of $y=x/N^{3/4}$. In the fluid phase, where the displacement $x$ is of order one, the minimum is located at $y=0$. On the other hand, in the condensed phase there is a single displacement with $x\sim O(N^{3/4})$, corresponding to $y>0$. Indeed, above the critical point $z_c$, the rate function $\psi_z(y)$ has two degenerate global minima, at $y=0$ and $y=y^*>0$. This is in agreement with the fact that $N-1$ displacements are of order one (corresponding to $y=0$) and one displacement is of order $N^{3/4}$ (corresponding to $y=y^*$). Finally, the fact that for $z_{\ell}<z<z_c$ a local minimum appears at $y=y^*>0$ means that the condensed phase is metastable for this range of parameters.

Since the condensate is a single large run (see Appendix \ref{app:one_cond}), the fluid fraction $m_f$, defined in Eq. \eqref{mf}, can be interpreted as the fraction of the total displacement $X$ associated to the other $N-1$ running phases, whose displacements $x_i$ are of order one. Indeed, using the expressions in Eqs. \eqref{mf} and \eqref{mc_expression}, it is possible to check that the condensate fraction $m_c$ and the fluid fraction $m_f$ sum to unity, i.e., that
\begin{equation}
m_c+m_f=1\,.
\label{sum_m}
\end{equation}
For $z<z_c$, we have shown that $m_f=1$ and $m_c=0$. Just above the critical point $z=z_c$, the condensate fraction $m_c$ jumps to the value $1/2$ and thus $m_f=1/2$. Increasing $z$ further, the value of $m_c$ increases while $m_f$ decreases. In the limit $z\to \infty$, we find that $m_c\to 1$ and $m_f\to 0$.

It is useful to denote by $X_{\rm fluid}=m_f X$ the total fluid displacement, i.e., the displacement associated with the $N-1$ runs which are not in the condensate. From Eq. \eqref{sum_m} we obtain
\begin{equation}
X_{\rm cond}+X_{\rm fluid}=X\,,
\end{equation}
where $X_{\rm cond}=m_c X$. Moreover, using Eq. \eqref{mc_limits}, it is easy to show that for large $z$
\be
X_{\rm fluid}\simeq \frac{2}{z^{1/3}}N^{3/4}\,,
\ee
meaning that, increasing the total displacement $X$ on a scale $N^{3/4}$, the fluid displacement $X_{\rm fluid}$ decreases.

This phenomenon is peculiar to this kind of first-order condensation transition. Indeed, for standard condensation transitions, as the ones observed in mass-transport models \cite{MEZ2005,EMZ06} or in other RTP models \cite{MLDM21}, the mechanism of condensation is quite different. In those models, for $X>X_c$ a condensate forms with $X_{\rm cond}=X-X_c$. In other words, all the excess displacement (or mass in the case of mass-transport models) above the threshold $X_c$ is entirely absorbed by the condensate. As a consequence, the displacement $X_{\rm fluid}$ due to the fluid phase freezes to the constant value $X_c$ for $X>X_c$. Thus, even if the fluid fraction $m_f=X_c/X$ decreases with $X$, the fluid displacement $X_{\rm fluid}$ remains constant. In the model considered in this article, if we increase $X$ by some amount $\Delta X=X-X_c$ above the critical value $X_c$, this additional displacement will be also absorbed by the condensate, as in the standard condensation described above. However, in addition to this increase, the condensate displacement will also snatch a part of the fluid displacement. Consequently, the increase in the condensate displacement $X_{\rm cond}$ will be larger than $\Delta X$ and the additional displacement is taken from the fluid displacement $X_{\rm fluid}$, which therefore decreases. Thus, in the case of a first-order condensation transition as studied here, the fluid and the condensed part of the trajectory are in some sense interacting even for $X>X_c$, i.e., when the system is fully in the condensed phase. This is at variance with condensation transitions of higher order where the fluid part of the trajectory is inert (frozen) for $X>X_c$.

\begin{figure}
\includegraphics[width=\columnwidth]{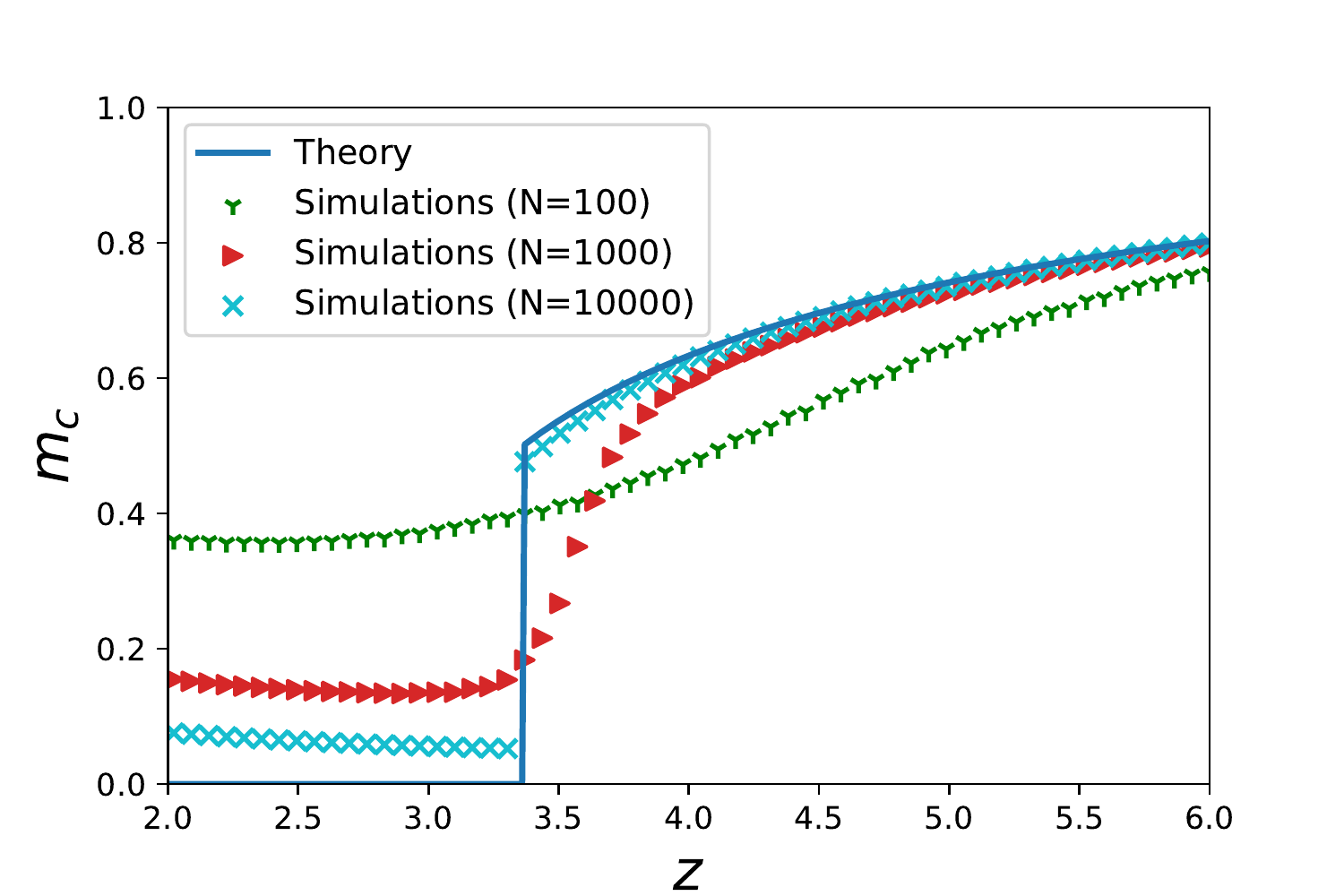}
\caption{The condensate fraction $m_c$ as a function of $z=X/N^{3/4}$. The continuous blue line corresponds to the exact result in Eq. \eqref{mc_expression}, while the symbols correspond to numerical simulations performed at different values of the number $N$ of running phases}
\label{fig:mc}
\end{figure}

Once we have identified the location $y=y^*$ of the condensate, it is relevant to investigate its shape. To do this, we need to expand the expression in Eq. \eqref{marg_x|X4} around $y=y^*$. This yields, after few steps of algebra,
\be 
p(x|X)\simeq p_{\rm cond}(x-y^* N^{3/4},N)\\
\label{marg_x|X5}
\ee
where
\be
p_{\rm cond}(y,N)\sim \exp\left[-\frac12 \psi''_z(y^*) \frac{y^2}{N}\right]\,.
\ee
Here $\psi''_z(y)$ denotes the second derivative of $\psi_z(y)$, given in Eq. \eqref{A_definition}, with respect to $y$. To obtain the result in Eq. \eqref{marg_x|X5}, we have used the fact that, above the transition, $\psi_z(y^*)=0$ and $\psi'_z(y^*)=0$. Overall, we have shown that, above the transition, a condensate appears at $x=y^* N^{3/4}$, with a Gaussian shape and standard deviation which grows as $\sqrt{N}$.

It is also possible to compute the area under the bump, which corresponds to the probability that a given site becomes the condensate. To do this, one needs to compute carefully the prefactor in Eq. \eqref{marg_x|X5}. We present the details of this computation in Appendix \ref{app:shape}, where we show that
\be
p_{\rm cond}(y,N)=\frac{1}{N^{3/2}}\sqrt{\frac{\psi''_z(y^*)}{\pi}} \exp\left[-\frac12 \psi''_z(y^*) \frac{y^2}{N}\right]\,.
\label{pcond_full}
\ee
From this expression, we obtain
\begin{equation}
\int_{-\infty}^{\infty}dy~p_{\rm cond}(y,N)=\frac{1}{N}\,,
\label{marg_normalization}
\end{equation}
meaning that above the transition the condensate is localized in just one of the $N$ sites. Note that it is possible to show that in the thermodynamic limit there can be at most one condensate (see Appendix \ref{app:one_cond}).

Finally, we introduce another order parameter, the participation ratio, which is often considered in the literature of condensation transitions \cite{GIL21,GIP21}. This quantity $Y_2(z)$ is defined as
\begin{equation}
Y_2(z)={\left\langle\frac{\sum_{i=1}^Nx_i^2}{\left(\sum_{i=1}^Nx_i\right)^2}\right\rangle}_z\,.
\label{Y2def}
\end{equation}
The notation $\langle\ldots\rangle_z$ denotes the average over the distribution of the displacements $x_1\,,\ldots\,,x_N$, conditioned on the value $X=z N^{3/4}$ of the total displacement. In the fluid phase, we expect the numerator in Eq. \eqref{Y2def} to scale as $O(N)$, since the terms $x_1\,,\ldots\,,x_N$ are of order one. On the other hand, the denominator is equal to $X^2\sim O(N^{3/2})$. Thus, for $z<z_c$ the participation ratio should vanish as $1/\sqrt{N}$. Conversely, in the condensed phase, a single variable $x_i$ absorbs a finite fraction of the displacement $X$. Thus, both numerator and denominator are expected to scale as $N^{3/2}$ and $Y_2(z)\sim O(1)$ in the large $N$ limit. For this reason, $Y_2(z)$ is a good order parameter for our system, where the corresponding control parameter is $z=X/N^{3/4}$.

Let us now compute the precise expression of $Y_2(z)$. First of all, using the fact that the variables $x_i$ are independent and that the denominator in Eq. \eqref{Y2def} is simply $X=zN^{3/4}$, we can rewrite the expression in Eq. \eqref{Y2def} as
\begin{equation}
Y_2(z)=\frac{ \langle x^2 \rangle_z}{z^2\sqrt{N}}\,,
\label{Y2comp}
\end{equation}
where 
\begin{equation}
\langle x^2\rangle_z=\int_{-\infty}^{\infty}dx~x^2~p(x|X)\,.
\end{equation}
For $z<z_c$, using the expression of $p(x|X)$ in Eq. \eqref{marg_x|X3}, it is easy to show that $\langle x^2\rangle_z\sim O(1)$ and thus that $Y_2(z)$ vanishes as $1/\sqrt{N}$ in the large-$N$ limit. Conversely, for $z>z_c$, we have shown that a bump appears in the tail of $p(x|X)$. It turns out that above the transition $\langle x^2\rangle_z$ is dominated by the contribution coming from this condensate bump. Indeed, using the expression of $p(x|X)$ in the vicinity of the bump, given in Eq. \eqref{pcond_full}, we obtain , to leading order,
\begin{equation}
\langle x^2\rangle_z\simeq {y^*}^2\sqrt{N}\,,
\end{equation}
where $y^*$ is given in Eq. \eqref{y_expression}. Pugging this expression in Eq. \eqref{Y2comp}, we obtain
\begin{equation}
Y_2(z)=\left(\frac{y^*}{z}\right)^2\,.
\end{equation}
Recalling the expression of the condensate fraction $m_c$ in Eq. \eqref{mc_expression}, we find that the participation ratio and the condensate fraction are simply related by
\begin{equation}
Y_2(z)=m_c^2\,.
\end{equation}
To summarize, we have shown that
\begin{equation}
Y_2(z)=\begin{cases}
0 & \text{for } z<z_c,\\
\\
\left[y^*(z)/z\right]^2 & \text{for } z>z_c\,,
\end{cases}
\end{equation}
where $y^*(z)$ is a given in Eq. \eqref{y_expression}. This exact result is shown in Fig. \eqref{fig:Y2} and is in good agreement with numerical simulations. Using the asymptic expressions of $m_c$ in Eq. \eqref{mc_limits}, we find 
\begin{equation}
Y_2(z)\simeq\begin{cases}
1/4+2^{-7/4}(z-z_c) & \text{for } z\to z_c^+,\\
\\
1-4 z^{-4/3} & \text{for } z\to\infty\,,
\end{cases}
\end{equation}
as given in Eq. \eqref{mc_lim}.

\section{Conclusions}

\label{sec:conclusions}

In this paper, we have investigated the position distribution of a single RTP moving in one dimension. We have assumed that the velocity of the particle during each running phase is independently drawn from a Gaussian distribution. We have computed the PDF $P(X,N)$ of the position $X$ of the particle after $N$ running phases, showing that this PDF is characterized by three different regimes in the limit of large $N$. In the typical regime $X\sim O(\sqrt{N})$ and the distribution $P(X,N)$ has a Gaussian shape, as predicted by the CLT. On the other hand, in the extreme large deviation regime $X\sim O(N)$, the PDF $P(X,N)$ has a stretched exponential form, signaling that the full displacement $X$ occurs in a single running phase. Finally, in the intermediate regime $X\sim O(N^{3/4})$, the PDF of $X$ is described by the rate function $F(z)$, where $z=X/N^{3/4}$. Below the critical value $z_c$, $F(z)$ is quadratic, meaning that $P(X,N)$ remains Gaussian up to $X=z_c N^{3/4}$, outside of the region predicted by the CLT. Interestingly, at the critical point $z_c$ the rate function $F(z)$ is non-analytic and for $z>z_c$ it is not quadratic anymore. In particular, it turns out that the first derivative of $F(z)$ is discontinuous at $z=z_c$, corresponding to a first-order dynamical phase transition. Note that this type of condensation transition is not present in the standard RTP model, where the velocity of the particle is constant.

We have provided a detailed analysis of the mechanism of the phase
transition. First of all, we have investigated the marginal
probability $p(x|X)$ of a single-run displacement, conditioned on the
total displacement $X$. We have shown that, above the transition, a
bump appears in the tail of $p(x|X)$, suggesting that the system
undergoes a condensation transition. In particular, we have shown that
above the transition a single running phase contributes to a
macroscopic fraction $m_c=X_{\rm cond}/X$ of $X$. We have observed that the mechanism of the transition is different from the one of standard condensation transitions, e.g., those observed in mass-transport models. Indeed, increasing the total displacement $X$ above the critical value $X_c$, the displacement $X_{\rm cond}$ contained in the condensate does not only absorb the excess displacement $\Delta X=X-X_c$, but it also snatches a part of the fluid displacement $X_{\rm fluid}$, i.e., the displacement associated to the remaining $N-1$ runs. Thus, $X_{\rm cond}>\Delta X$ in this case. This is different from the standard condensation transitions studied before, where $X_{\rm cond}=\Delta X$ and $X_{\rm fluid}$ remains frozen in the condensed phase. In addition, we have identified a relevant
order parameter associated to the phase transition: the participation
ratio $Y_2(z)$. We have shown that $Y_2(z)$ is zero below the
transition,  it is non-zero for $z>z_c$. We have
observed that this order parameter undergoes a jump $\Delta Y_2(z)=1/4$ exactly at $z=z_c$, in agreement with the fact that the transition
is first-order. Finally, we have performed extensive numerical
simulations to confirm our theoretical results. In order to study
numerically the probability of the rare events associated to the
large-deviation regime, we have employed a constrained Markov chain Monte Carlo algorithm.

We have shown that the problem investigated in this paper can be
mapped into the classical problem of finding the distribution of $N$
i.i.d. random variables. Despite the apparent simplicity of the setup,
we have observed that the PDF $P(X,N)$ is highly non-trivial in the
large $N$ limit and that a dynamical phase transition is observed
above a critical value of $X$. As mentioned in the introduction, an
alternative approach when studying the RTP model is to fix the total
elapsed time $T$, instead of the number $N$ of running phases. This
corresponds to the fixed-$T$ ensemble, which is usually harder to
treat analytically. Even if we expect the late-time behavior to be
similar for the fixed-$N$ and fixed-$T$ ensembles, for future works it
would be relevant to investigate the large deviation regime of
$P(X,T)$, i.e., the PDF of the RTP position $X$ after time $T$. It
would be interesting to compute the rate function associated to the
large-deviation regime $X\sim O(T^{3/4})$ and to recover the
first-order transition for this ensemble.

It is relevant to mention that if one considers a generic symmetric distribution $\mP(x)$ with
\begin{equation}
\mP(x)\sim e^{-a|x|^\beta}\,,\label{beta}
\end{equation}
with $a>0$ and $0<\beta<1$, the main results of this paper remain valid. Indeed, even if the scale at which the phase transition occurs is different [one can check that $X_c\sim O(N^{\alpha})$ with $\alpha=1/(2-\beta)$], the mechanism and the order of the transition remain the same. 

Moreover, it is interesting to notice that the criterion for condensation presented in Eq. \eqref{criterion} remains valid even if the motion of the particle between two tumbling events is not ballistic. Note that in this general case the expression of the distribution $\mP(x)$ of the displacement during a running phase would depend on the details of the dynamics. For instance, a condensation transition could be observed in RTP models which take into account rotational diffusion or fluctuation of the velocity within each running phase. Indeed, the presence of a condensation transition is guaranteed if (i) the displacements of the particle during different runs are i.i.d. random variables with distribution $\mP(x)$ and (ii) the PDF $\mP(x)$ satisfies the condition in Eq. \eqref{criterion}. Thus, the result in Eq. \eqref{criterion} appears to be quite robust and it is relevant to ask if condensation transitions as the one described in this work could be observed in experimental systems.

Finally, let us also mention that the problem of computing the distribution of the sum of several i.i.d. variables with a stretched exponential PDF appears in many other situations, including the problem of localization in the discrete nonlinear Schr\"odinger equation~\cite{GIL21,GIL21b}. Another example can be found in the context of one-dimensional Brownian motion with resetting \cite{EM11,EMS20}. Indeed, it is easy to show that the integral of the position of the Brownian particle between two resetting events will have a stretched-exponential distribution with $\beta=1/2$, where $\beta$ is the exponent defined in Eq. \eqref{beta}. For this reason, the integral of the position of the resetting Brownian motion can be written as the sum of several i.i.d. stretched-exponential variables and the system will display a first-order condensation transition at late times. Thus, we expect our results to be applicable to several problems besides the one considered in this paper.

\section*{Acknowledgments}

We thank P. Le Doussal, R. Livi and G. Schehr for useful discussions.

\appendix

\section{Asymptotic tails of $\mP(x)$}
\label{app:Asymp_Px}

In this Appendix we derive the asymptotic tails of $\mP(x)$ given in Eq. (\ref{x_marg.2a}). Let us focus on the case $x>0$ with $x\gg 1$. Performing the change of variable $\tau\to y=\tau/x^{2/3}$ in Eq. \eqref{x_marg.2a}, we obtain
\be 
\mP(x)=\frac{1}{\sqrt{2\pi}}\int_{0}^{\infty}dy\,\frac{1}{y}e^{-x^{2/3}[ y+1/(2y^2)]}\,.
\ee
For large $x$, this integral can be performed by saddle-point approximation, which yields
\be 
\mP(x)\simeq\frac{1}{\sqrt{3}x^{1/3}}e^{-(3/2) x^{2/3}}\,.
\ee
Repeating this argument in the case of $x<0$ with $|x|\gg 1$, we obtain
\be 
\mP(x)\simeq \frac{1}{\sqrt{3}|x|^{1/3}}e^{-(3/2)|x|^{2/3}}\,.
\ee
as given in Eq. \eqref{eq:mP-asymptotic}.

\section{Numerical simulations in the large-deviation regime}
\label{app:Monte-Carlo}

In this Appendix, we present the details of the numerical simulations that we have performed to confirm our analytical results. In order to study the large-deviation regime we employ a constrained MCMC algorithm, similar to the ones used in Refs. \cite{NMV10,NMV11,GM19,MLDM21}.

An RTP configuration with $N$ steps is described by the $N$ couples $\{(\tau_i,v_i)\}=\{(\tau_1,v_1)\,,\ldots(\tau_N,v_N)\}$. The probability weight associated to each configuration is
\begin{equation}
P(\{(\tau_i,v_i)\})=\prod_{i=1}^{N}p(\tau_i)W(v_i)\,,
\end{equation}
where $p(\tau)$ and $W(v)$ are given in Eqs. \eqref{duration_pdf} and \eqref{vel_pdf}.
The position $X$ of the particle after $N$ steps is given by
\begin{equation}
X=\sum_{i=1}^N \tau_i v_i\,.
\end{equation}
We want to estimate numerically the PDF $P(X,N)$ of the position $X$, in the limit of large $N$. If one is just interested in the typical fluctuations corresponding to $X\sim O(\sqrt{N})$, it is enough to employ a direct sampling strategy, drawing for each sample $N$ independent running times $\tau_1\,\ldots\tau_N$ from the PDF $p(\tau)$ and $N$ velocities $v_1\,,\ldots\,,v_N$ from $W(v)$. If one considers $10^{6}$ samples, this method allows to reach events that occur with probability of order $10^{-6}$ or higher. However, to sample the rare events corresponding to the large-deviation regime where $X\sim O(N^{3/4})$ one has to use a more sophisticated technique. Indeed, these events have probability smaller than $10^{-100}$ and direct sampling algorithms are not computationally feasible.

For this reason, we use a biased MCMC algorithm, which allows us to sample rare configurations which are characterized by an atypically large displacement $X$. First, we implement a MCMC dynamics in the space of configurations $\{(\tau_i,v_i)\}$, using the Metropolis-Hastings algorithm to guarantee that the RTP trajectories are sampled with the correct statistical weight. In particular, starting from any initial configuration, we choose the $i$-th running phase, where $i$ is a uniformly distributed random integer between $1$ and $N$, and we propose a move $(\tau_i,v_i)\to (\tau_i^{\rm new},v_i^{\rm new})$, where
\begin{equation}
\tau_i^{\rm new}=\tau_i+\delta\tau_i\,,
\end{equation}
and 
\begin{equation}
v_i^{\rm new}=v_i+\delta v_i\,.
\end{equation}
Here $\delta \tau_i$ and $\delta v_i$ are uniform random numbers in the intervals $(-a,a)$ and $(-b,b)$, respectively, where $a$ and $b$ are parameters of the algorithm. The move is accepted with probability
\begin{equation}
p_{\rm acc}=\min \left[ 1, \frac{p(\tau^i_{\rm new})W(v^i_{\rm new})}{p(\tau^i)W(v^i)}\right]\,,
\end{equation}
and rejected otherwise. Initially, we let the system evolve for $10^7$ sweeps, i.e., $10^7 N$ moves, in order to let the MCMC thermalize and then we measure the position $X$ of the RTP every $10^2$ sweeps, to avoid correlations. From these samples, we build an histogram that approximates the PDF $P(X,N)$. Up to this point, the MCMC algorithm is completely equivalent to the direct sampling strategy and it only allows to sample typical trajectories.

In order to investigate the large deviations of $P(X,N)$, we need to
bias the MCMC dynamics towards large values of $X$. For the sake of
simplicity we will focus on positive values of $X$, the case $X<0$ can
be treated analogously. We start by choosing some large value
$X^*$. Since we want to investigate the regime where $X\sim
O(N^{3/4})$, we will take $X^*\sim O(N^{3/4})$. We initialize the MCMC
from some initial condition with $X>X^*$. Then, we evolve the system
according to the MCMC dynamics described above, adding the hard
constraint $X>X^*$. In other words, attempted updates corresponding to $X<X^*$ are always rejected.

The histogram that we obtain from this biased algorithm will approximate the PDF $P(X,N|X>X^*)$, i.e., the PDF of $X$ conditioned on the event $X>X^*$. This quantity is then simply related to the PDF $P(X,N)$ by, for $X>X^*$,
\begin{equation}
P(X,N|X>X^*)=\frac{P(X,N)}{P(X>X^*)}\,.
\end{equation}
Taking the natural logarithm of both sides, we obtain
\begin{eqnarray}
&&\log\left[P(X,N|X>X^*)\right]\nonumber \\ &=&\log\left[P(X,N)\right]-\log\left[P(X>X^*)\right]\,.
\end{eqnarray}
Diving both sides by $\sqrt{N}$ and recalling that in the large-$N$ limit the rate function $F(z)$ is defined as
\begin{equation}
F\left(\frac{X}{N^{3/4}}\right)=-\frac{\log\left[P(X,N)\right]}{\sqrt{N}}\,,
\end{equation}
we find
\begin{equation}
F\left(\frac{X}{N^{3/4}}\right)=-\frac{\log\left[P(X,N|X>X^*)\right]}{\sqrt{N}}+C_{X^*}\,.
\end{equation}
In the equation above, we have defined the constant (with respect to $X$)
\begin{equation}
C_{X^*}=\frac{\log\left[P(X>X^*)\right]}{\sqrt{N}}\,.
\end{equation}

Thus, in order to estimate numerically $F(z)$ we need to compute the value of $C_{X^*}$. This can be achieved by the following strategy. First, we perform an unbiased simulation, which will allow us to estimate $F(z)$ in a small interval around the origin. Then, we choose a value of $X^*$ such that $z^*=X^*/N^{3/4}$ falls within the range of values for which $F(z)$ is known. The biased simulation will give us an estimate of $F(z)$ in a small region with $z>z^*$, up to the constant $C_{X^*}$. Since the two estimates of $F(z)$, the one obtained without the constraint and the one with the constraint, overlap for some values of $z$, one can compute the constant $C_{X^*}$ by matching the two curves. This allows us to know $F(z)$ in a slightly larger interval. Then, we continue by performing a new MCMC simulation with a larger value of $X^*$ and so on until $F(z)$ is known in a large-enough interval. Note that to speed up the algorithm the procedure above can be parallelized by choosing a fine enough grid of values $X^*$ in order to ensure the overlap between the different histograms. For instance, to obtain the numerical curves in Fig. \ref{fig:F} we have used $90$ equispaced values of $X^*$. 

Let us also mention that with the technique described above one can also obtain the marginal PDF $p(x|X)$ (see Fig. \ref{fig:marg}). From the MCMC dynamics one has access to the values of the single-run displacements $x_1\,,\ldots x_N$. Measuring a randomly chosen displacement at every step one can build an histogram which will approximate the PDF $p(x|X>X^*)$. However, it turns out that when $X^*$ is large, the system will typically stay in a small region to the right of $X^*$. In other words, even if $X$ is free to fluctuate during the simulation, it will typically remain close to $X^*$. Thus, one can approximate
\begin{equation}
p(x|X>X^*)\simeq p(x|X^*)\,.
\end{equation}
Alternatively, one can avoid this approximation by performing a MCMC dynamics at fixed $X$ (e.g., by proposing moves that conserve the total displacement $X$). However, the convergence of the algorithm turns out to be slower in this case. With the same technique one can also estimate $Y_2$ and $m_c$ as functions of $z$ (see Figs. \ref{fig:Y2} and \ref{fig:mc}). The results of our numerical simulations are in good agreement with the theory.

\section{Exact computation of the function $\chi(z)$}
\label{app:chi}

\begin{figure}
%%%
%%%
\centering
\includegraphics[scale=1]{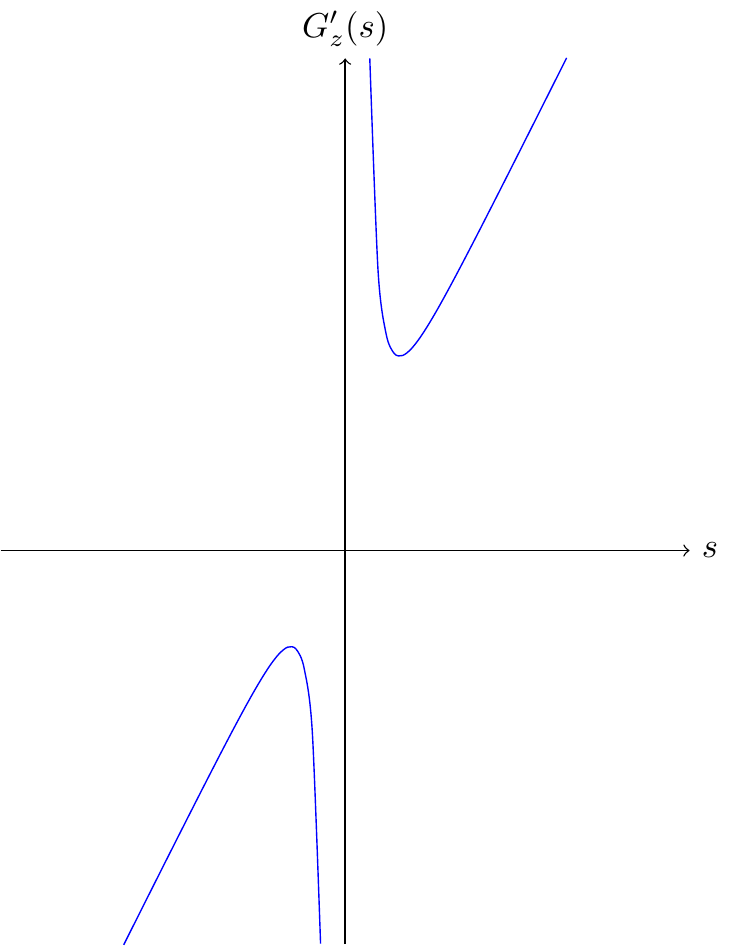} 
\caption{Plot of $G'_z(s)$, given in Eq. \eqref{SPE} as a function of $s$, for $z=1$. For this value of $z$ the saddle point condition in Eq. \eqref{SPE} has no real solutions.}
\label{fig:SPE}
\end{figure}

Our starting point is the integral in Eq. \eqref{anomalous_term2}, which reads
\bea
\label{anomalous_term4}
&&P_A(X= z N^{3/4},N)\\
&=&\frac{1}{\sqrt{2\pi} i} N
\int_{\Gamma^+} ds~\frac{1}{\sqrt{-s^2}} e^{\sqrt{N}G_z(s)}\,.\nonumber
\eea
where 
\be
G_z(s)=z s+s^2-\frac{1}{2s^2}\,.
\label{Gs}
\ee
We recall that we are considering the case $z>0$ and that the integral in Eq. \eqref{anomalous_term4} is performed over the contour $\Gamma^+$ in the complex $s$ plane, running in the negative real semiplane, parallel to the imaginary axis and with $\operatorname{Im}(s)>0$ and with $\operatorname{Re}(s)\to 0^-$ (see Fig. \ref{fig:Bromwich}). 

Our goal is to perform this integral by saddle-point approximation. To do that, let us consider the saddle-point equation
\be
G'_z(s)=z +2s+\frac{1}{s^3}=0\,,
\label{SPE}
\ee
where $G'_z(s)$ indicates the first derivative of $G_z(s)$ with respect to $s$. This function $G'_z(s)$ is shown in Fig. \ref{fig:SPE} as a function of $s$ for $z=1$. It is clear from Fig. \ref{fig:SPE} that Eq. \eqref{SPE} has no real solution for $z=1$. However, increasing $z$ amounts to a rigid upward vertical translation of the curve in Fig. \ref{fig:SPE}. Thus, at a critical value of $z$, that we call $z_{\ell}$, the $G_z'(s)$ curve will hit the negative $s$ axis and the saddle point equation \eqref{SPE} will have a single real solution $s^*<0$. Increasing $z$ further, Eq. \eqref{SPE} will have two real solutions, which we denote by $s_1$ and $s_2$, with $s_1<s_2<0$. It is easy to check that $s_1$ corresponds to a minimum of $G_z(s)$, while $s_2$ corresponds to a maximum.

Thus, for $z>z_{\ell}$ the complex integral in Eq. \eqref{anomalous_term4} can be computed as follows. First, we rotate the contour $\Gamma^+$ anticlockwise by an angle $\pi/2$, so that it now passes through the saddle points $s_1$ and $s_2$. Then, since $s_2$ is a maximum, the integral will be dominated, for large $N$, by contributions close to $s_2$ and one obtains
\bea
&&P_A(X= z N^{3/4},N)\simeq \frac{1}{-s_2 \sqrt{-G''_z(s_2)}} N^{3/4} e^{-\sqrt{N}\chi(z)}\,.\nonumber
\label{anomalous_term5}
\eea
where
\be 
\chi(z)=-G_z(s_2)\,,
\label{chi_definition}
\ee
and
\be
G''_{z_{\ell}}(s_2)=2-\frac{3}{{s_2}^4}<0\,,
\label{condition2}
\ee
where $G''_z(s)$ denotes the second derivative of $G_z(s)$ with respect to $s$.

We now need to compute the limit value $z_{\ell}$, which can be identified with the following argument. As explained above, at $z=z_{\ell}$, the saddle-point condition in Eq. \eqref{SPE} will have exactly one real solution $s^*<0$. Thus, $s^*$ satisfies the condition
\be
z_{\ell} +2s^*+\frac{1}{{s^*}^3}=0\,.
\label{condition1}
\ee
Moreover, it is clear from Fig. \ref{fig:SPE} that at $z=z_{\ell}$ the point $s^*$ corresponds to a maximum of the function $G'_z(s)$. Thus, one also has the condition
\be
G''_{z_{\ell}}(s^*)=2-\frac{3}{{s^*}^4}=0\,.
\label{condition_2}
\ee
Solving the two conditions in Eqs. \eqref{condition1} and \eqref{condition_2}, one finally finds that
\be
s^*=-\left(\frac32\right)^{1/4}
\label{s_star}
\ee
and
\be
z_{\ell}=4\left(\frac23\right)^{3/4}=2.95115\ldots
\ee
Then, for $z>z_{\ell}$ the saddle point $s_2$ is defined as the largest real root of Eq. \eqref{SPE} and can be computed by using Mathematica. Plugging this expression for $s_2$ into the definition of $\chi(z)$ in Eq. \eqref{chi_definition}, we find that
\begin{widetext}
\bea
&&\chi(z)=z^{2/3}
\left(\frac{z^{4/3}}{8}+\frac{1}{8} \left(\frac{64 \left(2/3\right)^{1/3}}{a(z)}+4 \left(2/3\right)^{2/3} a(z)\text{  }z^{4/3}+z^{8/3}\right)^{1/2}-\right.
\frac{1}{2} \left(-\frac{4 \left(2/3\right)^{1/3}}{a(z)}-\frac{a(z) z^{4/3}}{2\ 2^{1/3} 3^{2/3}}+\frac{z^{8/3}}{8}\right.\nonumber\\&+ &
\left.\left.\frac{z^4}{8 \sqrt{\frac{64 \left(2/3\right)^{1/3}}{a(z)}+4 \left(\frac{2}{3}\right)^{2/3} a(z)\text{  }z^{4/3}+z^{8/3}}}\right)^{1/2}\right)^{-2}+
\frac{z^{2/3}}{2} \left(-\frac{z^{4/3}}{8}-\frac{1}{8} \sqrt{\frac{64 \left(2/3\right)^{1/3}}{a(z)}+4 \left(2/3\right)^{2/3} a(z) z^{4/3}+z^{8/3}}\right.\nonumber\\ &+&
\frac{1}{2} \left(-\frac{4 \left(\frac{2}{3}\right)^{1/3}}{a(z)}-\frac{a(z) z^{4/3}}{2\ 2^{1/3} 3^{2/3}}+\frac{z^{8/3}}{8}+\right.
\left.\left.\frac{z^4}{8 \sqrt{\frac{64 \left(2/3\right)^{1/3}}{a(z)}+4 \left(2/3\right)^{2/3} a(z) z^{4/3}+z^{8/3}}}\right)^{1/2}\right)\,,
\label{chi_expression}
\eea

\end{widetext}
where
\be
a(z)=\left(9+\sqrt{3}\sqrt{27-\frac{2048}{z^2}}\right)^{1/3}\,.
\ee

\section{Asymptotics of $\chi(z)$}
\label{app:asymptotics}

In this section we want to compute the asymptotics of the function $\chi(z)$ at the edges of its domain $z_{\ell}<z<\infty$.

When $z\to z_{\ell}$ we already know that $s_2\to s^*$. Thus, plugging the value of $s^*$, given in Eq. \eqref{s_star}, into the definition of $\chi(z)$, given in Eq. \eqref{chi_definition}, we find that when $z\to z_{\ell}$
\be 
\chi(z)\to \sqrt{6}\,,
\ee
as given in Eq. \eqref{asymptotics_chi}.

In order to investigate the limit $z\to\infty$, it is useful to define the variable
\begin{equation}
\phi=-z^{1/3} s\,.
\label{phi_definition}
\end{equation}
Then the saddle point equation \eqref{SPE} can be rewritten in terms of $\phi$ as
\begin{equation}
2 z^{-4/3} \phi^4-\phi^3+1=0\,.
\label{SPE_phi}
\end{equation}
Note that after the transformation in Eq. \eqref{phi_definition}, the relevant saddle point is the smallest positive root $\phi_2$ of Eq. \eqref{SPE_phi}. Using the expression of the function $G_z(s)$, given in Eq. (\ref{Gs}), and the definition of $\chi(z)$, given in Eq. \eqref{chi_definition}, we obtain
\begin{equation}
\chi(z)=z^{2/3}\left(\frac12 \phi_2 +\frac{1}{{\phi_2}^2}\right).
\label{chi_phi}
\end{equation}

It turns out that Eq. \eqref{SPE_phi} is particularly useful to compute the large-$z$ asymptotics. Indeed, for large $z$, Eq. \eqref{SPE_phi} can be solved perturbatively and one obtains
\be 
\phi_2=1+\frac23 z^{-4/3}+o(z^{-4/3})\,.
\label{phi_asymptotics}
\ee
Plugging this expansion into Eq. \eqref{chi_phi} and expanding for large $z$, we obtain
\begin{equation}
\chi(z)=\frac32 z^{3/2}- z^{-2/3}+o(1)\,,
\end{equation}
as given in Eq. \eqref{asymptotics_chi}.

\section{Computation of the critical point $z_c$}
\label{app:zc}

In the main text we have shown that the rate function for the intermediate matching regime is given by [see Eq. \eqref{F}]
\be 
F(z)=\min\left[\frac{z^2}{4},\chi(z)\right]\,,
\ee
where $\chi(z)$ is given in Eq. \eqref{chi_expression}. It is easy to check, for instance numerically, that $\chi(z)>z^2/4$ for $z>z_c$, where $z_c$ is a constant of order one. Thus, one obtains 
\be
F(z)=
\begin{cases}
z^2/4 & \text{for } z<z_c,\\
\\
\chi(z) & \text{for } z>z_c.
\end{cases}
\ee
In this section, we want to compute exactly the critical point $z_c$, which is defined by the equation
\be 
\chi(z_c)=z_c^2/4\,.
\label{critical_equation}
\ee

Computing $z_c$ starting from the explicit expression of $\chi(z)$ in Eq. \eqref{chi_expression} appears to be rather challenging. However, using the representation of $\chi(z)$, given in Eq. \eqref{chi_phi}, in terms of the variable $\phi$, defined in Eq. \eqref{phi_definition}, solving Eq. \eqref{critical_equation} becomes simpler. First of all, from Eq. \eqref{SPE_phi}, evaluated at the critical point $z=z_c$, we find that
\be
z_c=\left(\frac{{\phi_2}^3-1}{2{\phi_2}^4}\right)^{-3/4}\,,
\label{cond1}
\ee
where we recall that $\phi_2$ is the smallest positive root of Eq. \eqref{SPE_phi}. Plugging the representation of $\chi(z)$ in terms of $\phi_2$, given in Eq. \eqref{chi_phi}, into Eq. \eqref{critical_equation}, we obtain
\be 
z_c^{2/3}\left(\frac12 \phi_2 +\frac{1}{{\phi_2}^2}\right)=z_c^2/4\,.
\label{cond2}
\ee
Finally, solving Eqs. \eqref{cond1} and \eqref{cond2}, it is easy to show that
\begin{equation}
z_c=2^{7/4}=3.36359\ldots
\end{equation}

\section{The number of condensates}

\label{app:one_cond}

In this appendix, we show that configurations with two condensates are less likely with respect to those with a single condensate. We start by rewriting the expression in Eq. \eqref{eq:PXN-free} as
\bea
&&P(X,N) =  \int_{-\infty}^{\infty}
dx_1~\mP(x_1)\int_{-\infty}^{\infty}
dx_2~\mP(x_2)\\
&\times&  \prod_{i=3}^N \int_{-\infty}^{\infty}
dx_i~\mP(x_i)~\delta\left(X-x_1-x_2-\sum_{i=3}^N x_i\right)\,,
\label{two_c}
\nonumber
\eea
where $\mP(X)$ is the single-run PDF, given in Eq. \eqref{x_marg.1}. Using again Eq. \eqref{eq:PXN-free}, we can write Eq. \eqref{two_c} as
\bea
P(X,N) &=&  \int_{-\infty}^{\infty}
dx_1~\mP(x_1)\int_{-\infty}^{\infty}
dx_2~\mP(x_2)\nonumber\\
&\times & P(X-x_1-x_2,N-2)\,.
\eea
In the regime $X\sim O(N^{3/4})$ we use the large deviation form of $P(X,N)$, given in Eq. \eqref{regimes}, and the large-$x$ behavior of $\mP(x)$, given in Eq. \eqref{eq:mP-asymptotic}, and we obtain
\begin{eqnarray}
&& e^{-\sqrt{N}F(X/N^{3/4})}\sim \int_{-\infty}^{\infty}dx_1~\int_{-\infty}^{\infty}dx_2~\\
&\times & e^{-\sqrt{N}F[(X-x_1-x_2)/N^{3/4}]-(3/2)|x_1|^{2/3}-(3/2)|x_2|^{2/3}}\,.\nonumber
\end{eqnarray}
Note that we are using the large-$x$ asymptotics of $\mP(x)$ because we are probing for configurations where $x_1$ and $x_2$ represent two condensates and hence are of order $O(N^{3/4})$.
Using the scaled variables $z=X/N^{3/4}$, $y_1=x_1/N^{3/4}$, and $y_2=x_2/N^{3/4}$, we obtain the relation
\begin{eqnarray}
&& e^{-\sqrt{N}F(z)}\sim \int_{-\infty}^{\infty}dy_1~\int_{-\infty}^{\infty}dy_2~\\
&\times & e^{-\sqrt{N}\left\{F[(z-y_1-y_2)/N^{3/4}]+(3/2)|y_1|^{2/3}+(3/2)|y_2|^{2/3}\right\}}\,.\nonumber
\end{eqnarray}
The variables $y_1$ and $y_2$ represent the fraction of the total displacement contained in the two condensates. Thus, the presence of two condensate would correspond to both $y_1>0$ and $y_2>0$. Performing both integrals via saddle-point approximation, we find
\begin{eqnarray}
&&F(z)\\
&=&\min_{0\leq y_1\leq z}~\min_{0\leq y_2\leq z-y_1}
\left[F(z-y_1-y_2)+\frac32 y_1^{2/3}+\frac32 y_2^{2/3}\right]\,.\nonumber
\end{eqnarray}
Performing the change of variables $(y_1,y_2)\to (y_1,y=y_1+y_2)$, we obtain
\begin{eqnarray}
&&F(z)\\
&=&\min_{0\leq y\leq z}~\min_{0\leq y_1\leq y}
\left[F(z-y)+\frac32 y_1^{2/3}+\frac32 (y-y_1)^{2/3}\right]\,,\nonumber
\end{eqnarray}
which can be rewritten as
\begin{eqnarray}
&&F(z)\\
&=&\min_{0\leq y\leq z}~
\left[F(z-y)+\min_{0\leq y_1\leq y} \left(\frac32 y_1^{2/3}+\frac32 (y-y_1)^{2/3}\right)\right]\,.\nonumber
\end{eqnarray}
It is easy to show that the minimum over $y_1$ is obtained for $y_1=0$ and $y_2=y>0$, or by symmetry for $y_2=0$ and $y_1=y>0$. Recalling that the presence of two condensates would require both $y_1>0$ and $y_2>0$, we have shown that configurations with two condensates cannot be observed in this model when $N\to \infty$.

Note that we have also verified that there can be only one condensate by showing that [see Eq. \eqref{area}]
\begin{equation}
\int_{-\infty}^{\infty}dy~p_{\rm cond}(y,N)=\frac{1}{N}\,,
\end{equation}
i.e., that the area under the condensation bump is $1/N$. This is in agreement with the fact that only one of the $N$ running phases can become the condensate.

\section{Equivalence with the result of Ref. \cite{BKL20}}
\label{app:equivalence}

\begin{figure}
\includegraphics[width=\columnwidth]{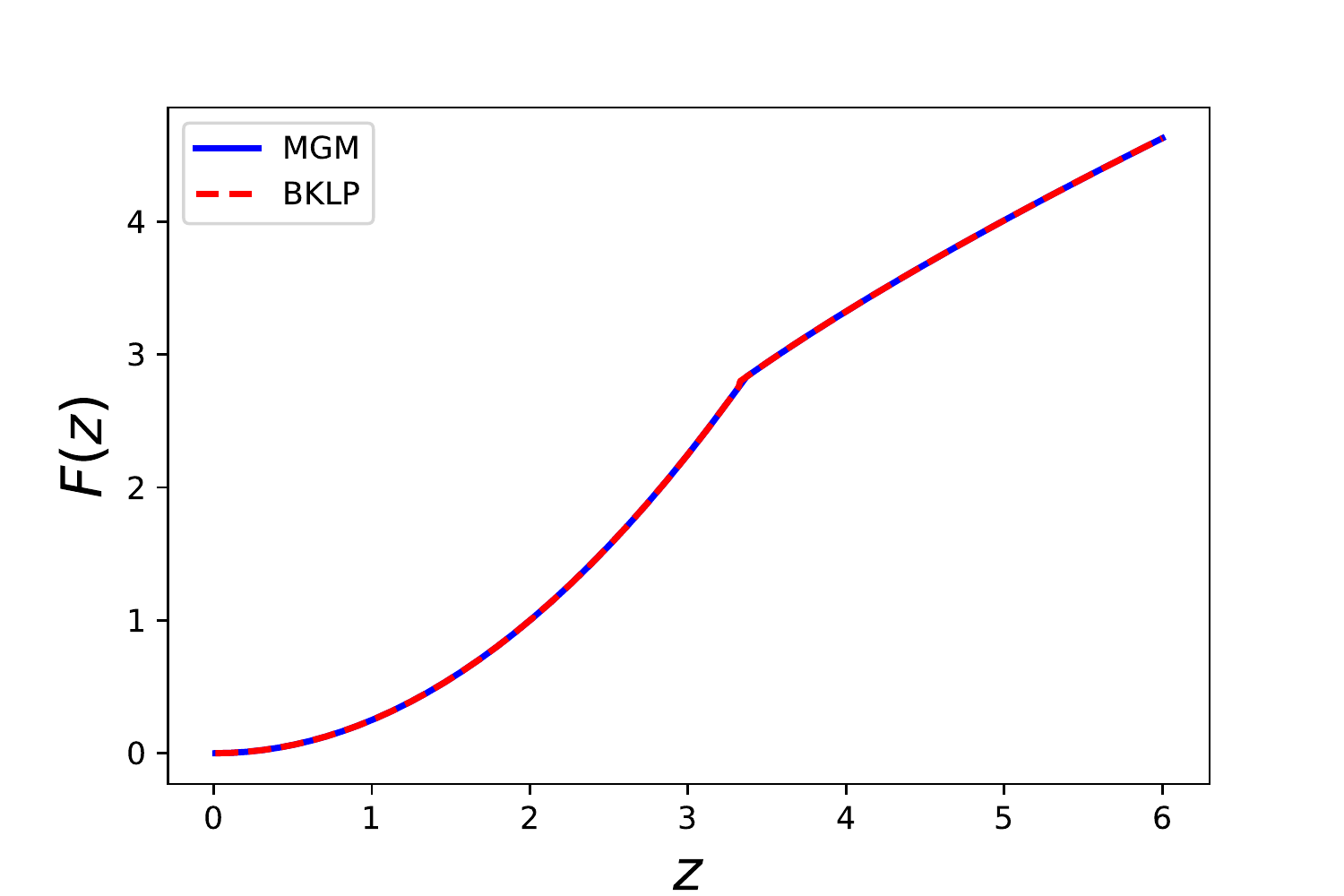}
\caption{Comparison between the rate function $F(z)$ computed by Brosset et al. (BKLP), given in Eq. \eqref{F_brosset}, and our result (MGM), given in Eq. \eqref{F_intro}.}
\label{fig:comparison}
\end{figure}

In Ref. \cite{BKL20} Brosset et al. derived by mathematical methods the rate function $F(z)$, defined in Eq. \eqref{definiton_F}, showing that
\begin{equation}
F(z)=\min_{0\leq y\leq z}\left[\frac{(z-y)^2}{4}+\frac{3}{2}y^{2/3}\right]\,.
\label{F_brosset}
\end{equation}
At first sight, this result seems different from the one derived in this paper and given in Eq. \eqref{F_intro}. However, performing the minimization numerically, we find that they are equivalent [see Fig. \ref{fig:comparison}]. In this appendix we show analytically that the two expressions are equivalent.

Let us define
\begin{equation}
g_z(y)=\frac{(z-y)^2}{4}+\frac{3}{2}y^{2/3}\,,
\label{gz_def}
\end{equation}
so that we can write the expression in Eq. \eqref{F_brosset} as
\begin{equation}
F(z)=\min_{0\leq y\leq z}g_z(y)
\end{equation}
Thus, we look for the solution $y^*$ of the equation
\begin{equation}
g'_z(y^*)=\frac{y^*-z}{2}+ {y^*}^{-1/3}=0\,.
\label{min_condition}
\end{equation}
Moreover, since $y^*$ is a minimum, it also satisfies
\begin{equation}
g''_z(y^*)=-\frac13 {y^*}^{-4/3}+\frac12>0\,.
\label{min_condition2}
\end{equation}
Note that the minimum of $g_z(y)$ could be also reached at $y=0$ or $y=z$. 

One can show that $y^*$ is related to the solution $\phi_2$ of the saddle point equation \eqref{SPE_phi} by the relation
\begin{equation}
y^*=z\phi_2^{-3}\,.
\label{yphi2}
\end{equation}
Note that both $y^*$ and $\phi_2$ depend on $z$. Thus, from Appendix \ref{app:asymptotics}, we know that Eq. \eqref{min_condition} will have no solution for $z<z_{\ell}$, implying that the minimum will be reached either at $y=0$ or at $y=z$. It is easy to show that, for $z<z_{\ell}$, $g_z(0)<g_z(z)$ and therefore we obtain 
\begin{equation}
F(z)=g_z(0)=\frac{z^2}{4}\,,
\end{equation}
for $z<z_{\ell}$. In the region $z>z_{\ell}$ a solution $y^*$ of Eq. \eqref{min_condition} that satisfy the condition \eqref{min_condition2} exists. However, it is easy to check that just above $z_{\ell}$ the global minimum is still reached at $y=0$. Increasing $z$, the value of the minimum corresponding to $y^*$ decreases until at some value $z=z_c$, the global minimum is reached at $y=y^*$. Thus, the critical point $z=z_c$ is defined as
\begin{equation}
g_{z_c}(0)=g_{z_c}(y^*)\,,
\label{cond_zc}
\end{equation}
where $y^*$ is the solution of Eq. \eqref{min_condition} that satisfies the condition \eqref{min_condition2}. One can check that the point $y=z$ is never the global minimum.

Plugging the expression of $g_z(y)$, given in Eq. \eqref{gz_def}, into Eq. \eqref{cond_zc}, we find that $z_c$ satisfies
\begin{equation}
\frac{z_c^2}{4}=\frac{(z_c-y^*)^2}{4}+\frac{3}{2}{y^*}^{2/3}\,.
\end{equation}
Using Eq. \eqref{yphi2}, we get
\begin{equation}
\frac{z_c^2}{4}=\frac{(z_c-z_c\phi_2^{-3})^2}{4}+\frac{3}{2}\left(z_c\phi_2^{-3}\right)^{2/3}\,,
\label{equal1}
\end{equation}
where we recall that $\phi_2$ is the smallest positive solution of the saddle point equation \eqref{SPE_phi}. Moreover, from Eq. \eqref{SPE_phi}, we obtain
\begin{equation}
z^{4/3}=2\frac{\phi_2^4}{\phi_2^3-1}\,.
\label{equal2}
\end{equation}
Eq. \eqref{equal1} can be rewritten as
\begin{equation}
\frac{z_c^{4/3}}{4}\left[1-(1-\phi_2^{-3})^2\right]=\frac32 \phi_2^{-2}\,.
\end{equation}
Plugging the expression for $z_c^{4/3}$ given in Eq. \eqref{equal2} and solving for $\phi_2$ yields
\begin{equation}
\phi_2=2^{1/3}\,,
\end{equation}
for $z=z_c$. Using Eq. \eqref{equal2}, we obtain $z_c=2^{7/4}$. Thus, for $z<z_c$ we have shown that
\begin{equation}
F(z)=\frac{z^2}{4}\,,
\end{equation}
in agreement with our result in Eq. \eqref{F_intro}.

Finally, we need to show that for $z>z_c$ the two results in Eqs. \eqref{F_brosset} and \eqref{F_intro} coincide, i.e., that
\begin{equation}
\frac{(z-y^*)^2}{4}+\frac{3}{2}{y^*}^{2/3}=\chi(z)\,,
\label{ychi}
\end{equation}
where $\chi(z)$ is given in Eq. \eqref{chi_expression}. From Eq. \eqref{chi_phi} we know that $\chi(z)$ can be written in terms of the variable $\phi_2$ as
\begin{equation}
\chi(z)=z^{2/3}\left(\frac12 \phi_2 +\frac{1}{{\phi_2}^2}\right).
\label{chip}
\end{equation}
The left-hand side of Eq. \eqref{ychi} can be rewritten, using Eq. \eqref{yphi2}, as
\begin{equation}
\frac{(z-y^*)^2}{4}+\frac{3}{2}{y^*}^{2/3}=z^{2/3}\left[\frac{z^{4/3}}{4}(1-\phi_2^{-3})^2+\frac32 \phi_2^{-2}\right]\,.
\end{equation}
Using Eq. \eqref{equal2}, we rewrite the term $z^{4/3}$ as
\begin{equation}
\frac{(z-y^*)^2}{4}+\frac{3}{2}{y^*}^{2/3}=z^{2/3}\left[\frac{\phi_2^4(1-\phi_2^{-3})^2}{2(\phi_2^3-1)}+\frac32 \phi_2^{-2}\right]\,,
\end{equation}
which can be rewritten, after few steps of algebra, as
\begin{equation}
\frac{(z-y^*)^2}{4}+\frac{3}{2}{y^*}^{2/3}=z^{2/3}\left(\frac12 \phi_2 +\frac{1}{{\phi_2}^2}\right)\,.
\end{equation}
Recalling Eq. \eqref{chip}, we finally obtain the result in Eq. \eqref{ychi}. Thus, we have shown that for any $z>0$ the two representations of $F(z)$ in Eq. \eqref{F_intro} and \eqref{F_brosset} are equivalent.

Finally, we provide a physical interpretation of the formula in Eq. \eqref{F_brosset}. In the rest of this appendix we will only consider the case $X>0$. We start from the definition of the PDF of $X$, given in Eq. \eqref{eq:PXN-free},
\be
P(X,N) =  \prod_{i=1}^N \int_{-\infty}^{\infty}
dx_i~\mP(x_i)~\delta\left(X-\sum_{i=1}^N x_i\right),
\label{PXNapp}
\ee
where $\mP(X)$ is the single-run PDF, given in Eq. \eqref{x_marg.1}. The expression in Eq. \eqref{PXNapp}, can be rewritten as
\bea
&&P(X,N) =  \int_{-\infty}^{\infty}
dx_1~\mP(x_1)\\
&\times&  \prod_{i=2}^N \int_{-\infty}^{\infty}
dx_i~\mP(x_i)~\delta\left(X-x_1-\sum_{i=2}^N x_i\right)\,.\nonumber
\eea
Using Eq. \eqref{PXNapp}, we obtain
\begin{equation}
P(X,N) =  \int_{-\infty}^{\infty}
dx_1~\mP(x_1)P(X-x_1,N-1)\,.
\label{PXNapp2}
\end{equation}
In the condensed phase, the term $\mP(x_1)$ represents the condensate while the term $P(X-x_1,N-1)$ is associated with the other $N-1$ variables which are in the fluid phase, meaning that their contribution is of order one. Note that we have shown that no more than one condensate can appear in the thermodynamic limit (see App. \ref{app:one_cond}).

We now focus in the regime where $X\sim O(N^{3/4})$. Plugging the large deviation form of $P(X,N)$ in the fluid-phase, given in Eqs. \eqref{regimes} and \eqref{F_intro}, and the large-$x$ behavior of $\mP(x)$, given in Eq. \eqref{eq:mP-asymptotic}, into Eq. \eqref{PXNapp2}, we find 
\begin{equation}
e^{-\sqrt{N}F(X/N^{3/4})}\sim \int_{-\infty}^{\infty}dx_1~e^{-(X-x_1)^2/(4N)-(3/2)|x_1|^{2/3}}\,.
\end{equation}
Using the scaled variables $y=x_1/N^{3/4}$ and $z=X/N^{3/4}$, we obtain
\begin{equation}
e^{-\sqrt{N}F(z)}\sim \int_{-\infty}^{\infty}dy~e^{-\sqrt{N}(z-y)^2/4-(3/2)\sqrt{N}|y|^{2/3}}\,.
\end{equation}
For large $N$, one can perform the integral over $y$ via saddle-point approximation, yielding
\be 
F(z)=\inf_{-\infty<y<\infty}\left[\frac{(z-y)^2}{4}+\frac{3}{2}|y|^{2/3}\right]\,.
\ee
Moreover, it is easy to show that configurations with $y<0$ or $y>z$ are never optimal. Thus, we obtain
\be 
F(z)=\min_{0\leq y\leq z}\left[\frac{(z-y)^2}{4}+\frac{3}{2}y^{2/3}\right]\,,
\label{relation_app}
\ee
in agreement with Eq. \eqref{F_brosset}.

\section{Location of the condensate}
\label{app:location}

In this Appendix, we derive an exact expression for the location $y^*$ of the condensate, presented in Section \ref{sec:marginal}. For $z>z_c$, we recall that $m_c=y^*/z$ is the fraction of the total displacement belonging to the condensate, where $z=X N^{-3/4}$. The variable $y^*$ is defined as
\begin{equation}
y^*=\operatorname{argmin}_{y>0}\left[\psi_z(y)\right]\,.
\label{y*_def_app}
\end{equation}
where
\begin{equation}
\psi_z(y)=\frac{3}{2}y^{2/3}+F(z-y)-F(z)\,,
\label{A_definition_app}
\end{equation}
and $F(z)$ is given in Eq. \eqref{F_cases}. We limit our discussion to the case $z>0$, the complementary case $z<0$ can be obtained by symmetry.

From Eq. \eqref{y*_def_app} we find that $y^*$ satisfies the equation
\be 
\psi'_z(y^*)=0\,,
\label{cond_y*}
\ee 
where $\psi'_z(y)$ denotes the first derivative of $\psi_z(y)$ with respect to $y$. We now assume that $y^*>z-z_c$ (to be verified a posteriori). Under this assumption the condition in Eq. \eqref{cond_y*} becomes
\begin{equation}
2{y^*}^{-1/3}+y^*-z=0\,.
\end{equation}
Since $y^*$ has to be a minimum, one also has the additional condition
\be 
\psi''_z(y^*)=-\frac13 {y^*}^{-4/3}+\frac12>0\,,
\label{cond2_y*}
\ee 
where $\psi''_z(y)$ denotes the second derivative of $\psi_z(y)$ with respect to $y$.

Remarkably, it turns out that the solution $y^*$ of the conditions above can be exactly related to the solution $\phi_2$ of the saddle-point equation \eqref{SPE_phi} as
\begin{equation}
y^*=z~ {\phi_2}^{-3}\,.
\label{y*_phi}
\end{equation}
Thus, from the results in Appendix \ref{app:chi}, we know that no solution of the saddle-point equation exists for $z<z_{\ell}=4~(2/3)^{3/4}$. Thus, for $z<z_{\ell}$ the function $\psi_z(y)$ will not have any minimum for $y>0$ and no condensation is possible. For $z>z_{\ell}$ a minimum of the function $\psi_z(y)$ appears at $y=y^*>0$. However, it is easy to check that, for $z_{\ell}<z<z_c$, this minimum will correspond to a non-zero value of the function $\psi_z(y^*)$ and thus the probability density of configurations with $y>0$ will decay exponentially fast with $N$ [see Eq. \eqref{marg_x|X4}]. Finally, above the transition $\psi_z(y^*)$ becomes, using Eq. \eqref{F_cases},
\be 
\psi_z(y^*)=\frac32 {y^*}^{2/3}+\frac{(z-y^*)^2}{4}-\chi(z)\,,
\ee
where $\chi(z)$ is defined in Eq. \eqref{chi_definition}. We recall that the function $\chi(z)$ can be expressed in terms of the variable $\phi_2$, as given in Eq. \eqref{chi_phi}. Using also the representation of $y^*$ in terms of $\phi_2$, given in Eq. \eqref{y*_phi}, we obtain
\begin{eqnarray}
&&\psi_z(y^*)\\
&=&\frac32 {z}^{2/3} \phi_2^{-2}+z^2 \frac{(1-{\phi_2}^{-3})^2}{4}-z^{2/3}\left(\frac12 \phi_2 +{\phi_2}^{-2}\right)\,. \nonumber
\label{A_y*}
\end{eqnarray} 
From Eq. \eqref{SPE_phi}, we obtain
\begin{equation}
z^{4/3}=\frac{2{\phi_2}^{4}}{{\phi_2}^3-1}\,.
\label{z^4/3}
\end{equation}
Thus, writing the term $z^2$ in Eq. \eqref{A_y*} as $z^2=z^{2/3}~z^{4/3}$ and using Eq. \eqref{z^4/3}, we obtain, after few steps of algebra, that for $z>z_c$ (see also Fig. \ref{fig:A})
\begin{equation}
\psi_z(y^*)=0\,.
\end{equation}
Thus, for $z>z_c$ a condensate appears at $x=y^*N^{3/4}$ [see Eq. \eqref{marg_x|X4}].

It is also possible to find an explicit expression for $y^*$. Solving the conditions in Eqs. \eqref{cond_y*} and \eqref{cond2_y*}, we obtain
\begin{widetext}
\bea
&&y^*=z
\left(\frac{z^{4/3}}{8}+\frac{1}{8} \left(\frac{64 \left(2/3\right)^{1/3}}{a(z)}+4 \left(2/3\right)^{2/3} a(z)\text{  }z^{4/3}+z^{8/3}\right)^{1/2}-\right.
\frac{1}{2} \left(-\frac{4 \left(2/3\right)^{1/3}}{a(z)}-\frac{a(z) z^{4/3}}{2\ 2^{1/3} 3^{2/3}}+\frac{z^{8/3}}{8}\right.\nonumber\\&+ &
\left.\left.\frac{z^4}{8 \sqrt{\frac{64 \left(2/3\right)^{1/3}}{a(z)}+4 \left(\frac{2}{3}\right)^{2/3} a(z)\text{  }z^{4/3}+z^{8/3}}}\right)^{1/2}\right)^{-3}\,,
\label{y_expression}
\eea
\end{widetext}
where
\be
a(z)=\left(9+\sqrt{3}\sqrt{27-\frac{2048}{z^2}}\right)^{1/3}\,.
\ee
From this expression in Eq. \eqref{y_expression}, we obtain, for $z=z_c=2^{7/4}$, $y^*=2^{3/4}$ and hence that $m_c=y^*/z=1/2$, as given in the first line of Eq. \eqref{mc_limits}. It is easy to show that $y^*$ is an increasing function of $z$ and that our assumption $y^*>z-z_c$ is always satisfied. Moreover, close to $z=z_c$, we find that
\begin{equation}
y*\simeq 2^{3/4}+\frac32 (z-z_c)\,,
\end{equation} 
and thus that
\begin{equation}
m_c\simeq\frac12+\frac{z-z_c}{2^{7/4}}\,.
\end{equation}

Finally, to investigate the large-$z$ behavior of $y^*$, we use express $y^*$ in terms of the variable $\phi_2$, using Eq. \eqref{y*_phi}. Using the asymptotics of $\phi_2$ for large $z$, given in Eq. \eqref{phi_asymptotics}, we obtain
\begin{equation}
y^*\simeq z-2 z^{-1/3}\,,
\end{equation}
and, using $m_c=y^*/z$, we finally find that, for large $z$
\begin{equation}
m_c\simeq 1-2 z^{-4/3}\,,
\end{equation}
as given in the second line of Eq. \eqref{mc_limits}.

\section{Shape of the condensate}
\label{app:shape}

In Section \ref{sec:marginal} we have shown that for $z>z_c$ the condensate appears as a bump in the marginal probability $p(x|X)$ of the single displacement $x$ conditioned on the total displacement $X$. We have also shown that the condensate is located in the proximity of $x=y^* N^{3/4}$, where the function $p(x|X)$ can be approximated as
\bea
&&p(x|X)\simeq p_{\rm cond}(x-y^*N^{3/4},N) \\
&\sim &\exp\left[-\frac12 \psi''_z(y^*) \frac{(x-y^* N^{3/4})^2}{N}\right]\,,\nonumber
\eea
where $y^*$ is given in Eq. \eqref{y_expression} and $\psi''_z(y^*)$ is given in Eq. \eqref{cond2_y*}. The goal of this appendix is to compute the full distribution $p_{\rm cond}(y,N)$ (and not only the exponential part given above). Knowing the full expression of the function $p_{\rm cond}(y,N)$ is useful because the integral 
\be
\int_{-\infty}^{\infty}dy~p_{\rm cond}(y,N)
\ee
is the probability that a specific displacement among $x_1\,,\ldots\,,x_N$ becomes the condensate. In the presence of a single condensate one expects this probability to be $1/N$.

Our starting point is the exact expression in Eq. \eqref{marg_x|X}, which reads
\bea
p(x|X)&=&\mP(x)\frac{P(X-x,N-1)}{P(X,N)}\,.
\label{marg_x|X_app}
\eea
We are interested in the regime where $x=yN^{3/4}$ and $X=zN^{3/4}$, the variables $y$ and $z$ being of order one.
Note that the function $P(X,N)$ has different expressions below and above the transition. Since $z>z_c$ the term $P(X,N)$ is given by [see Eq. \eqref{anomalous_term5}]
\bea
&&P(X,N)\simeq \frac{1}{-s_2 \sqrt{-G''_z(s_2)}} N^{3/4} e^{-\sqrt{N}G_z(s_2)}\,.
\label{anomalous_app}
\eea
where
\be 
G_z(s_2)=z~ s_2+{s_2}^2-\frac{1}{2{s_2}^2}\,,
\label{G_app}
\ee
and
\be
G''_{z_{\ell}}(s_2)=2-\frac{3}{{s_2}^4}<0\,.
\label{G''_app}
\ee
We recall that the variable $s_2$ is the largest negative solution of the saddle point equation \eqref{SPE} and that it is related to the variable $\phi_2$ by $s_2=-z^{-1/3}\phi_2$.

In Appendix \ref{app:location} we have shown that the condensate is located at $y=y^*$ and that $y^*>z-z_c$, thus $P(X-x,N-1)$ is given by the Gaussian weight (see Eq. \eqref{gaussian_term_2})
\be
P(X-x,N-1)\simeq\frac{1}{2\sqrt{\pi N}}e^{-\sqrt{N} (z-y)^2/4}\,.
\label{gaussian_app}
\ee
Finally, we can approximate $\mP(x)$, using the expansion in Eq. \eqref{eq:mP-asymptotic}, as
\be 
\label{mp_yy}
\mP(x=yN^{3/4}) \approx \frac{N^{-1/4} }{\sqrt{3}\,
   y^{1/3}}\, e^{-\frac{3}{2}\, \sqrt{N} y^{2/3}}\, .
\ee
Thus, plugging the expressions in Eqs. \eqref{anomalous_app}, \eqref{gaussian_app}, and \eqref{mp_yy}, into Eq. \eqref{marg_x|X_app}, we obtain 
\bea
&&p(x|X)\simeq \frac{ - s_2 \sqrt{-G''_z(s_2)}}{\sqrt{3\pi} y^{1/3}}\frac{1}{N^{3/2}}\\
&\times &\exp\left[-\sqrt{N}\left(\frac{3}{2}y^{2/3}+(z-y)^2/4-G_z(s_2)\right)\right]\,.\nonumber
\label{marg_x|X_app2}
\eea

To investigate the shape of the condensate, we now expand the expression \eqref{marg_x|X_app2} around $y^*$, the location of the condensate. Thus, setting $x=N^{3/4}(y^*+w N^{-1/4})$ (with $w$ of order one) and expanding for large $N$ we obtain
\bea
\label{marg_x|X_app3}
p(x|X) \simeq p_{\rm cond}(\sqrt{N}w,N)&=&\frac{ -  ~s_2 \sqrt{-G''_z(s_2)}}{2\sqrt{3\pi} {y^*}^{1/3}}\\ &\times &\frac{1}{N^{3/2}}\exp\left[-\frac{1}{2} \psi''_z(y^*) w^2\right]\,,\nonumber
\eea
where $\psi''_z(y)$ is given in Eq. \eqref{cond2_y*}.

We can now compute the integral of the condensate distribution $p_{\rm cond}(x,N)$. Using the expression in Eq. \eqref{marg_x|X_app3} with $w=N^{-1/2}(x-N^{3/4}y)$ we obtain
\be 
\int_{-\infty}^{\infty}dx\,p_{\rm cond}(x,N)=\frac{1}{\sqrt{6}}\frac{ - s_2 \sqrt{-G''_z(s_2)}}{ {y^*}^{1/3}\sqrt{\psi''_z(y^*)}}\frac{1}{N}\,.
\ee
It is useful to write the right-hand side of this equationd in terms of the variable $\phi_2$, defined in Appendix \ref{app:chi} as the smallest positive root of Eq. \eqref{SPE_phi}. Indeed, using the relations $y^*=z {\phi_2}^{-3}$ and $s_2=-z^{-1/3}\phi_2$, we obtain
\be 
\int_{-\infty}^{\infty}dx\,p_{\rm cond}(x,N)=\frac{1}{\sqrt{6}}z^{-2/3}{\phi_2}^2\frac{   \sqrt{-G''_z(-z^{-1/3}\phi_2)}}{\sqrt{\psi''_z(z {\phi_2}^{-3})}}\frac{1}{N}\,.
\ee
Using the expressions for $G''_z(s)$ and $\psi''_z(s)$, given in Eqs. \eqref{G''_app} and \eqref{cond2_y*}, we obtain
\be 
\int_{-\infty}^{\infty}dx\,p_{\rm cond}(x,N)=\frac{1}{\sqrt{6}}\frac{   \sqrt{3 -2 z^{-4/3}{\phi_2}^{4}}}{\sqrt{1/2-z^{-4/3} {\phi_2}^{4}/3}}\frac{1}{N}=\frac{1}{N}\,,
\ee
as anticipated in Eq. \eqref{marg_normalization}. This result implies that the condensate is localized in just one of the displacements $x_1\,,\ldots\,,x_N$.

\end{document}